%% file: geomlag.tex
\documentclass[fleqn]{article}

\usepackage{amsmath,amssymb,amsthm}
\usepackage{bm}
\usepackage{color}
\usepackage{pifont,txfonts}
\usepackage{natbib}

\ifx\pdftexversion\undefined
  \usepackage[dvips]{graphics}
\else
  \usepackage[pdftex]{graphics}
\fi

\input{defs.tex}\allowdisplaybreaks

\theoremstyle{plain}
\newtheorem{theorem}{Theorem}
\newtheorem{proposition}[theorem]{Proposition}
\newtheorem{lemma}[theorem]{Lemma}
\newtheorem{corollary}[theorem]{Corollary}

\newtheoremstyle{slantbody}{\topsep}{\topsep}{\slshape\def\defemph{\defemphrm}}{}{\bfseries}{.}{6pt}{\thmname{#1}\thmnumber{ #2}\thmnote{ #3}}
\theoremstyle{slantbody}
\newtheorem{definition}{Definition}

\theoremstyle{remark}
\newtheorem{example}{Example}
\newtheorem{remark}{Remark}

\newcommand{\elsqed}{}

\begin{document}

\title{Geometric Discrete Analogues of Tangent~Bundles and
  Constrained~Lagrangian~Systems}

\author{
Charles Cuell and George W.\ Patrick$^\dag$\\[.05in]
\small\it Applied Mathematics and Mathematical Physics\\[-.05in]
\small\it Department of Mathematics and Statistics\\[-.05in]
\small\it University of Saskatchewan\\[-.05in]
\small\it Saskatoon, Saskatchewan, S7N~5E6, Canada
}

\date{\small June 2008$^\ddag$}

\maketitle

\renewcommand{\thefootnote}{}
\footnotetext{$^\dag$Funded by the Natural Sciences and Engineering Reseach Council, Canada}
\footnotetext{$^\ddag\backslash\mbox{today}$: \today}

\vspace*{-.3in}
\begin{abstract}\noindent\input{geomlag-abstract.tex}\end{abstract}

\input{geomlag-contents.tex}

\footnotesize

\input{geomlag.bbl}
\end{document}

%% file: defs.tex
\newcommand{\defemphrm}[1]{{\rm #1}}
\def\defemph{\defemphsl}

\newcommand{\ann}{\operatorname{ann}}
\newcommand{\vrt}{\operatorname{vert}}
\newcommand{\hor}{\operatorname{hor}}
\newcommand{\fdim}{\operatorname{fdim}}
\newcommand{\cofdim}{\operatorname{cofdim}}
\newcommand{\CoAd}{\operatorname{CoAd}}
\newcommand{\LLdA}{\mbox{\rm LdA}}
\newcommand{\LdA}{\LLdA}
\newcommand{\LdAp}{\LLdA$^\prime$}
\newcommand{\LdApd}{\LLdA$^\prime_{\mathrm d}$}
\def\kC{{\mathcal C}}
\def\kD{{\mathcal D}}
\def\kE{{\mathcal E}}
\def\kF{{\mathcal F}}
\def\kG{{\mathcal G}}
\def\kK{{\mathcal K}}
\def\kL{{\mathcal L}}
\def\kM{{\mathcal M}}
\def\kN{{\mathcal N}}
\def\kP{{\mathcal P}}
\def\kQ{{\mathcal Q}}
\def\kV{{\mathcal V}}
\def\kW{{\mathcal W}}
\def\RR{{\mathbb R}}
\def\fg{{\mathfrak g}}

\newcommand{\w}[1]{#1}
\newcommand{\dbt}[1]{\ifcase#1{}\or.5\or1\or1.5\or2\or2.5\or3\or3.5\or4\or4.5\fi}
\newcommand{\+}[1]{\mskip\dbt#1mu}
\renewcommand{\-}[1]{\mskip-\dbt#1mu}
\renewcommand\![1]{{\bm{{#1}}}}

\def\ClAuDiAeatblanks{\global\@ignoretrue}
\def\ClAuDiAstop{\relax}
\def\ClAuDiAsplit #1#2\stop{\string #1}
\def\[#1\]{\protect{\if\ClAuDiAsplit #1\stop[%
{\ClAuDiAwithlabel #1\ClAuDiAstop}\else%
{\ClAuDiAwithoutlabel #1\ClAuDiAstop}\fi}\ignorespaces}%
\def\ClAuDiAwithlabel[#1]#2\ClAuDiAstop{\protect{\begin{equation}\label{#1}\begin{split}#2\end{split}\end{equation}}}
\def\ClAuDiAwithoutlabel#1\ClAuDiAstop{\protect{\begin{equation*}\begin{split}#1\end{split}\end{equation*}}}

\newcommand{\set}[2]{\bigl\{\mskip1mu #1:#2\mskip1mu\bigr\}}
\newcommand{\sset}[1]{\bigl\{\mskip1mu#1\mskip1mu\bigr\}}

\let\macron=\=
\def\={\ifmmode\equiv\else \macron\fi}

\let\temp=\colon\def\colon{\temp\mathopen{}}

\def\mathrlap{\mathpalette\mathrlapinternal}
\def\mathrlapinternal#1#2{\rlap{$\mathsurround=0pt#1{#2}$}}

%% file: geomlag-abstract.tex
Discretizing variational principles, as opposed to discretizing differential equations, leads to discrete-time analogues of mechanics, and, systematically, to geometric numerical integrators. The phase space of such variational discretizations is often the set of configuration pairs, analogously corresponding to initial and terminal points of a tangent vectors. We develop alternative discrete analogues of tangent bundles, by extending tangent vectors to finite curve segments, one curve segment for each tangent vector. Towards flexible, high order numerical integrators, we use these discrete tangent bundles as phase spaces for discretizations of the variational principles of Lagrangian systems, up to the generality of nonholonomic mechanical systems with nonlinear constraints. We obtain a self-contained and transparent development, where regularity, equations of motion, symmetry and momentum, and structure preservation, all have natural expressions. 

%% file: geomlag-contents.tex
%
\section{Introduction}
%
%
\noindent A \defemph{discretization} of a Lagrangian system $L\colon
\!T\kQ\rightarrow\RR$ consists of
\begin{enumerate}
  \item
    a \defemph{time step} $h>0$;
  \item
    the \defemph{discrete phase space} $\kQ\times\kQ=\sset{(q^+,q^-)}$,
    thought of as a discrete version of the tangent bundle $\!T\kQ$;
  \item
    a \defemph{discrete Lagrangian}
    $L_h\colon\kQ\times\kQ\rightarrow\RR$, obtained by approximately
    integrating $L$ over an appropriate interpolation from $q^-$ to
    $q^+$.
\end{enumerate}
The discrete Lagrangian $L_h$ and the discrete phase space
$\kQ\times\kQ$ together define a \defemph{discrete Lagrangian system.}
\defemph{Evolutions} are sequences $q_k\in\kQ$, $k=1,\ldots,N$ that
are critical points of the \defemph{discrete action} $S$, defined by
\[
  S\=\sum_{k=1}^NL_h(q_k,q_{k-1}),
\]
subject to the constraint $q_0$ and $q_N$ constant. As is easily
shown, $q_k$ is an evolution if and only if it satisfies the
\defemph{discrete Euler-Lagrange equation}
\[[eq:MV-DEL]
  \frac{\partial L_h}{\partial q^+}(q_k,q_{k-1})
    +\frac{\partial L_h}{\partial q^-}(q_{k+1},q_k)=0.
\]
Lagrangian  discretizations lead to (implicit, symplectic, and momentum
conserving) numerical methods because Equation~\eqref{eq:MV-DEL} can
be used to advance through time $h$ by stepping states
$(q_k,q_{k-1})$ to $(q_{k+1},q_k)$.

Such Lagrangian discretizations are towards discrete Lagrangian models
that reflect physical reality so well that they have a stature with
continuous Lagrangian models. States of a continuous model~---~such as
values of the independent variables of a differential equation~---~and
states of a discrete model~---~such as the pairs of configurations
used in the map implied by~\eqref{eq:MV-DEL}, can both serve as
abstract representations of system states. Most important is not the
particular representation of the states, but  whether the states
evolve as the physical system does. If errors, in either a discrete or
a continuous model, are below measurement errors of a physical system,
and neither the discrete nor the continuous model violates fundamental
physical principles to that accuracy, then both models have a similar
stature.  Lagrangian discretizations also provide variational discrete
analogues of continuous Lagrangian systems. They are of interest in
themselves, and they provide a framework for the analysis,
understanding, and development of geometric integration
algorithms~\cite{HairerE-LubichC-WannerG-2006-1} for Lagrangian
systems as purely mathematical objects. For more details,
see~\cite{CortesJ-MartinezS-2001-1,
  deLeonM-deDiegoD-SantamariaMerinoA-2004-1,
  FedorovYN-ZenkovDV-2005-1, MarsdenJE-WestM-2001-1,
  McLachlanR-PerlmutterM-2006-1, WendlandtJM-MarsdenJE-1997-1}.

In this article, we further develop discretizations of Lagrangian
systems. We refer to the discretizations outlined above
as \defemph{Moser-Veselov~(MV) discretizations}. Our discrete
Lagrangian systems replace the MV discrete phase space $\kQ\times\kQ$
with a discrete phase space~$\kV$ consisting of curve segments in
$\kQ$ which one-to-one correspond with elements of $\!T\kQ$. To any
such discrete Lagrangian system, there is naturally associated a
isomorphic MV discrete Lagrangian system, obtained by identifying our
curve segments with their boundaries. Viewing tangent bundles as
curve segments is generally consistent with viewing discretizations in
general as attaching to a manifold finite rather than infinitesimal
objects~\cite{BobenkoAI-SurisYB-2005-1}.

Systematically using curve segments provides theoretical flexibility
and geometric clarity. For example, our curve segments can be
naturally shrunk, and this helps to analyze limits where the time step
tends to zero. The interpolating curves of MV discretizations are
obtained implicitly from boundary value problems, with boundary values
the two configurations of the MV~discrete states. We finesse this
implicit dependence by directly using the interpolating curves. We
achieve a self contained variational theory, which does not depend on
discrete versions of the Legendre transform, nor on any canonical
formalism on the cotangent bundle. We show that the entire development
extends to nonholonomic systems with nonlinear constraints. We extend
the curvature conditions for holonomic subsystems of nonholonomic
systems~\cite{PatrickGW-2007-1} to the discrete case, and prove the
nonholonomic momentum
equation~\cite{BlochAM-KrishnaprasadPS-MarsdenJE-MurrayRM-1996-1,
               CortesJ-MartinezS-2001-1}
in our context. As well, we show how
our discrete Lagrangian systems specialize to discrete
\emph{holonomic} systems.

Some notations: Unless otherwise noted, objects are sufficiently
smooth to permit the required operations. If $\kM$ is a manifold and
$v_m,w_m\in\w{\!T\-2\kM}$, then define
\[
  \w{\vrt_{v_q}\-3}w_q\equiv\frac d{dt}\biggr|_{t=0}(v_q+tw_q).
\]
If $\pi\colon E\rightarrow\kM$ is a vector bundle, and
$z\in\w{\!T\-4_}{0_m}E$ i.e.~if $z$ is a tangent vector at the zero
section, then we denote the horizontal and vertical parts of $z$ by
$\hor z\in\w{\!T\-4_m\kM}$ and $\vrt z\in E_m$, respectively. We
denote the fiber dimension of a fiber bundle by~$\fdim$ and the fiber
codimension of a subbundle by~$\cofdim$.  To reduce double subscripts,
we sometimes use the functional notation $x(k)$ instead of $x_k$ for a
sequence. If $A$ is a set, then we will use the notation $A[M,N]$ for
the sequences in $v(k)\in A$, $k=M,\ldots,N$.  If $\kG$ acts smoothly
on a manifold $\kM$ then we denote the Lie algebra of $\kG$ by $\fg$,
and the \defemph{infinitesimal generator of $\xi\in\fg$ at $m\in\kM$}
by
\[
  \xi m\=\frac{d}{dt}\biggr|_{t=0}\exp(\xi t)m.
\]
Assembling these into  a vector  field gives $\xi_\kM(m)\=\xi m$.

%
\section{Discretizations of Tangent Bundles}
\label{sec:discretizations}
%
%
\begin{figure}[ht]
  \begin{center}\input{fig4_pstex_t.tex}\end{center}
\end{figure}
\noindent Let $\kM$ be a manifold and $m\in\kM$. Two curves $c\colon
(a,b)\rightarrow\kM$ and $\tilde c\colon(\tilde a,\tilde b)\rightarrow
\kM$ with $0\in(a,b)$ and $0\in(\tilde a,\tilde b)$ are
\defemph{tangent at $m$} if 1)~$c(0)=\tilde c(0)=m$; and 2)~$\phi\,
c(t)-\phi\,\tilde c(t)=O(t^2)$ in any chart $\phi$ of $\kM$ with
domain including $m$. Tangency at $m$ is an equivalence relation, and
the \defemph{tangent space} $\w{\!T\-4_m\kM}$ at $m\in\kM$ may be
defined~\cite{AbrahamR-MarsdenJE-RatiuTS-1988-1} as the set of
equivalence classes of curves at $m$. Our discretizations of
Lagrangian systems depend on the development of a discretization of a
tangent bundle $\w{\!T\-2\kM}$ as assignments of curve segments in
$\kM$ to tangent vectors of $\kM$. We will require a parameter $h$
such that $\w{\!T\-2\kM}$ is obtained in the limit
$h\rightarrow0^+$. So, we posit a map $\psi(h,t,m)$, with values in
$\kM$, and obtain the curve segments $t\mapsto\psi(h,t,m)$:

\begin{definition}\label{def:discretization-of-tangent-bundle}
  A \defemph{$C^k$ discretization of $\w{\!T\-2\kM}$}, $k\ge1$, is a tuple
  $(\psi,\alpha^+,\alpha^-)$, where
  \[
    \psi\colon U\subseteq\RR^2\times\kM\rightarrow\kM,\qquad
    \alpha^+\colon[0,a)\rightarrow\RR_{\ge0},\qquad
    \alpha^-\colon[0,a)\rightarrow\RR_{\le0},
  \]
  are such that
  \begin{enumerate}
    \item
      $\psi$ is continuous, $U$ is open, and
      $\sset{0}\times\sset{0}\times\kM\subseteq U$;
    \item
      $\alpha^+,\alpha^-$ are $C^1$, and $\alpha^+(h)-\alpha^-(h)=h$;
    \item
      $\psi(h,0,v_m)=m$, and
      $\displaystyle\frac{\partial\psi}{\partial t}(h,0,v_m)=v_m$;
    \item\label{enum:def-boundary-maps}
      the \defemph{boundary maps} defined by
      \[[100]
        \partial^-_h(v_m)\=\psi\bigl(h,\alpha^-(h),v_m\bigr),\qquad
        \partial^+_h(v_m)\=\psi\bigl(h,\alpha^+(h),v_m\big),
      \]
      are $C^k$ in $(h,v_m)$, and
      \[[101]
        \frac d{dh}\biggr|_{h=0}\partial^+_h(v_m)=\dot\alpha^+v_m,\qquad
        \frac d{dh}\biggr|_{h=0}\partial^-_h(v_m)=\dot\alpha^-v_m,
      \]
      where
      \[
        \dot\alpha^+\=\frac{d\alpha^{\mbox{}\mathrlap{+}}}{dh}(0),\qquad
        \dot\alpha^-\=\frac{d\alpha^{\mbox{}\mathrlap{-}}}{dh}(0).
      \]
  \end{enumerate}
\end{definition}

\begin{remark}
  Putting $h=0$ in $\alpha^+(h)-\alpha^-(h)=h$ gives
  $\alpha^+(0)=\alpha^-(0)=0$ because $\alpha^+\ge0$ and
  $\alpha^-\le0$. Differentiating $\alpha^+(h)-\alpha^-(h)=h$  at
  $h=0$ gives $\dot\alpha^+-\dot\alpha^-=1$. If $\psi$ is $C^1$, then
  Assumptions~\eqref{101} are superfluous, since
  \[
    \frac d{dh}\biggr|_{h=0}\partial^+_h(v_m)
      =\frac d{dh}\biggr|_{h=0}\psi\bigl(h,\alpha^+(h),v_m\bigr)
      =\frac d{dh}\biggr|_{h=0}\psi\bigl(h,0,v_m\bigr)
        +\frac d{dh}\biggr|_{h=0}\psi\bigl(0,\alpha^+(h),v_m\bigr)
      =\dot\alpha^+v_m,
  \]
  and similarly with $\partial^-_h$. The definition allows $\psi$ to
  be only piecewise smooth in $h,t$.
\end{remark}

For all $v_m\in\kM$, the set $\set{(h,t)}{(h,t,v_m)\in U}$ is open and
contains $h=0,t=0$, and so contains the set
$\set{(h,t)}{h\in[0,b),\,\alpha^-(h)\le t\le\alpha^+(h)}$ for some
$0<b<a$. So assigned to every $v_m\in\w{\!T\-2\kM}$ and small enough
$h>0$ is the curve segment
\[
  t\mapsto\psi(h,t,v_m),\qquad\alpha^-(h)\le t\le\alpha^+(h),
\]
which, since $\psi(h,0,v_m)=m$ and
$\displaystyle\frac{\partial\psi}{\partial t}(h,0,v_m)=v_m$, is a
curve at $m$ which is tangent to $v_m$. \emph{A discretization
$(\psi,\alpha^+,\alpha^-)$ assigns to every $v_m$ a curve segment that
can be thought of as a translational step like $hv_m$.}

\begin{example}\label{Example-trivial-discretization}
  Let $\kM\=\RR^N$, $0\le\gamma\le 1$, $\alpha^+(h)\=\gamma h$,
  $\alpha^-(h)\=-(1-\gamma)h$, and $\psi(t,h,v_m)\=m+tv$. More
  generally, let $X$ be any second order vector field on $\kM$,
  $0\le\gamma\le 1$ and define $\psi^X(h,t,v_q)\=\w{\tau\-2_\kM}\+2
  \w{F^X_{\-6t}}(v_q)$ where $\w{F^X_{\-6t}}$ is the flow of $X$ and
  $\w{\tau\-2_\kM}\colon\w{\!T\-2\kM}\to\kM$ is the canonical
  projection. $\psi^X$ and $\alpha^+,\alpha^-$ is the
  \defemph{$X$-discretization with bias $(\alpha^-,\alpha^+)$}.
\end{example}

\begin{example}
  Let $X$ be a vector field on a manifold $\kM$. A \defemph{one-step
  numerical method} for $X$ is a map $\varphi\colon
  U\subseteq[0,\infty)\times\kM\rightarrow\kM$ such that
  \begin{enumerate}
    \item $\sset{0}\times\kM\subseteq U$;
    \item $\varphi(0,m)=m$ for all $m\in\kM$;
    \item $\displaystyle\frac d{dt}\biggr|_{t=0}\varphi(t,m)=X(m)$.
  \end{enumerate}
  If $\varphi$ is a one-step numerical method for a second order
  vector field on $\w{\!T\-2\kM}$ then
  $\psi(h,t,v_q)\=\w{\tau\-2_\kM}\+4\varphi(t,v_q)$ is a
  discretization of $\w{\!T\-2\kM}$.
\end{example}

Generally, when we speak of a \defemph{discretization} we mean
a \emph{family of discrete objects} parametrized by $h\in\RR$, such
that a continuous target is approached as $h\to0$. A \defemph{discrete
object} is an \emph{instance of a discretization}, obtained by fixing
$h$ to a particular value and possibly dropping data not required to
make operational the discrete representation of the continuous
target. For tangent bundles, we choose the transition from
discretization to discrete as the juncture at which we drop the curve
segments in discretizations of tangent bundles, retaining only their
endpoints:

\begin{definition}\label{def:discrete-tangent-bundle}
  Let $\kM$ be a manifold. A \defemph{discrete tangent bundle of $\kM$}
  is a tuple $(\kV,\partial^+,\partial^-)$, where $\kV$ is a manifold,
  $\dim\kV= 2\dim\kM$ and $\partial^+\-2\colon\kV\rightarrow\kM$ and
  $\partial^-\-2\colon\kV\rightarrow\kM$ satisfy
  \begin{enumerate}
    \item\label{enum:discrete-tangent-bundle-1}
      $\partial^+$ and $\partial^-$ are submersions such that $\ker
      \!T\partial^+\cap\ker\!T\partial^-=0$; and
    \item\label{enum:discrete-tangent-bundle-2}
      for all $m\in\kM$, the \defemph{backward fiber}
      $\kV^+_{\-5m}\=(\partial^+)^{-1}(m)$ and the \defemph{forward fiber}
      $\w{\kV^-_{\-5m}}\=(\partial^-)^{-1}(m)$ meet in exactly one point,
      denoted $0_m$.
  \end{enumerate}
  The \defemph{discrete zero section} is
  $0_\kV\=(\partial^\pm)^{-1}\Delta(\kM\times\kM)$, where
  $\Delta(\kM\times\kM)$ is the diagonal of $\kM\times\kM$.
\end{definition}

\begin{remark}\label{rem:partial-pm-diffeo}
  Let $\partial^\pm\colon\kV\rightarrow\kM\times\kM$ be defined by
  $\partial^\pm(p)\=\bigl(\partial^+(p),\partial^-(p)\bigr)$.
  Item~\ref{enum:discrete-tangent-bundle-1} of
  Definition~\ref{def:discrete-tangent-bundle} implies that
  $\w{\!T\-4_}v\partial^\pm$ is a linear isomorphism for all
  $v\in\kV$, and therefore $\partial^\pm$ is a local
  diffeomorphism. Also, Item~\ref{enum:discrete-tangent-bundle-2} of
  Definition~\ref{def:discrete-tangent-bundle} implies that
  $\partial^\pm$ is bijective from $0_\kV$ to $\Delta(\kM\times\kM)$
  so that the local diffeomorphism $\partial^\pm$ is a diffeomorphism
  of $0_\kV$ to $\Delta(\kM\times\kM)$. $0_\kV$ is a closed submanifold
  of $\kV$ because $\Delta(\kM\times\kM)$ is a closed submanifold of
  $\kM\times\kM$, and Theorem~1 (semiglobal inverse function theorem)
  of~\cite{CuellC-PatrickGW-2007-1} provides open neighborhoods $U$ of
  $0_\kV$ and $V$ of $\Delta(\kM\times\kM)$ for which $\partial^\pm$
  is a diffeomorphism.
\end{remark}

\begin{remark}
  Our discrete tangent bundles are similar to the groupoid based
  constructions of discrete phase spaces for Lagrangian systems
  in~\cite{WeinsteinA-1996-1}. Indeed, to any discrete tangent bundle
  $\kV$ there is an associated Lie groupoid consisting of 1)~sequences
  $v_k$ in $\kV$ which satisfy $\partial^+(v_k)=\partial^-(v_{k+1})$,
  and the reverses of these, 2)~units the elements $0_m$, and
  3)~source and target maps $\partial^+$ and $\partial^-$. Some of the
  constructions below are the same as those found in the groupoid
  context. One can regard discrete tangent bundles of $\kM$ to be
  groupoids over $\kM$ for which the set of irreducible elements $\kV$
  is a manifold which satisfies
  Item~\ref{enum:discrete-tangent-bundle-1} of
  Definition~\ref{def:discrete-tangent-bundle}. To the extent of this
  article, the algebraic structure of the groupoid seems to generate
  more ambiguity than it does clarity. For example, starting as in
  Definition~\ref{def:discrete-tangent-bundle} with a tangent vector
  $v\in\kV$, one might include the formal reverse of $v$ in order to
  have its groupoid inverse. There will generally be another,
  different, element in $\kV$ with the same source and target as that
  formal reverse. And, the product of that with the original $v$ is a
  two element sequence that starts and ends at the same place of
  $\kM$, but it is not the same as the discrete zero vector.
\end{remark}

\begin{remark}
  Given any manifold $\kM$, the tuple
  $(\kM\times\kM,\w{\partial^+_{\-2\kM\times\kM}},\w{\partial^-_{\-2\kM\times\kM}})$
  is a discrete tangent bundle, where
  \[
    \kM\times\kM\=\sset{(m^+,m^-)},\qquad
    \w{\partial^-_{\-2\kM\times\kM}}(m^+,m^-)\=m^-,\qquad
    \w{\partial^+_{\-2\kM\times\kM}}(m^+,m^-)\=m^+.
  \]
  Thus the usual MV discretizations of the tangent bundle are special
  cases of Definition~\ref{def:discrete-tangent-bundle}. For a discrete
  tangent bundle $(\kV,\partial^+,\partial^-)$, the map $\partial^\pm$
  is typically a diffeomorphism in the region of interest, so one can
  in principle, using $\partial^\pm$,
  replace any discrete tangent
  bundle $(\kM,\partial^+,\partial^-)$ as in
  Definition~\ref{def:discrete-tangent-bundle} by its image in $\kM\times\kM$,
  dispense with the maps $\partial^+$ and $\partial^-$, and use the
  corresponding MV discrete tangent bundle:
  $$
  \includegraphics{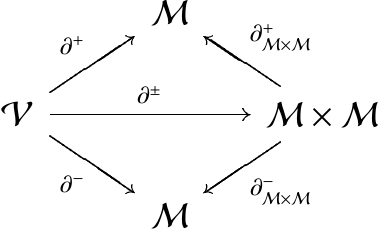}
  $$
  However, the freedom of including $\partial^+$ and $\partial^-$ in
  the definition of a discrete tangent bundle, and the abstraction of
  the discrete tangent vectors as elements of a manifold $\kV$, is
  helpful and clarifying.
\end{remark}

Since \defemph{discretizations} are to be families of
\defemph{discrete analogues}, it is necessary to show that a
discretization of a tangent bundle gives discrete tangent bundles for
sufficiently small $h$
(Proposition~\ref{prp:discretization-gives-discrete} below). This is
not immediate because there is a singularity at $h=0$. To see the
problem, consider the example $\kM\=\RR^N$, with discretization
$\psi\bigl(h,t,(m,v)\bigr)\=m+tv+O(h^2)$. To show
Item~\ref{enum:discrete-tangent-bundle-2} of
Definition~\ref{def:discrete-tangent-bundle}, it is sufficient to show
the $\partial_h^\pm$ is a diffeomorphism near $h=0$ i.e.\ that the
equations
\[[eq:example-mpm-eqs]
  m^-=m+\alpha^-(h)v+O(h^2),\qquad
  m^+=m+\alpha^+(h)v+O(h^2),
\]
can be solved uniquely and smoothly for $(m,v)$ in terms of
$(m^+,m^-)$. Because there is no a~priori knowledge of the details of
the $O(h^2)$ term, the proof of that has to be perturbative from
$h=0$. However, when $h=0$, Equations~\eqref{eq:example-mpm-eqs}
cannot be solved at for $v$ because they reduce to $m^-=m$, $m^+=m$.
Replacing $v$ with $\tilde v\=v/h$ would solve the problem but would
require, at later stages, smoothness assumptions in $v/h$, and would
make the subsequent development awkward and less general. But,
replacing $(m^+,m^-)$ with the new variables $(\bar m,z)$ defined by
\[[eq:example-blowup]
  \bar m\=\frac{m^++m^-}2,\qquad z\=\frac{m^+-m^-}h,
\]
converts Equations~\eqref{eq:example-mpm-eqs} to
\[[eq:example-mpm-xeqs]
  \bar m=m+\frac{\alpha^+(h)+\alpha^-(h)}2v+O(h^2),\qquad
  z=v+O(h).
\]
We remark that a division by $h$ has reduced the order
of the trailing term in the second equation. At $h=0$,
Equations~\eqref{eq:example-mpm-xeqs} are
\[
  \bar m=m,\qquad
  z=v,
\]
which has solution $m=\bar m$, $v=z$. By the implicit function
theorem, Equation~\eqref{eq:example-mpm-xeqs} may be solved for $m,v$
in terms of $\bar m,z$ for sufficiently small $h$, and hence, through
Equation~\eqref{eq:example-blowup}, for $m,v$ in terms of $(m^+,m^-)$,
as required. To obtain a result near $h=0$ which is valid for $m,v$ near
the whole zero section $0(\w{\!T\-2\kM})$ i.e.\ local along $h$ but global
along $0(\w{\!T\-2\kM})$, we again make use of Theorem~1 (semiglobal inverse
function theorem) of~\cite{CuellC-PatrickGW-2007-1}.

\begin{proposition}\label{prp:discretization-gives-discrete}
  Let $(\psi,\alpha^+,\alpha^-)$ be a discretization of the tangent
  bundle of $\kM$ and let $\kM_0\subset\kM$ be a relatively compact
  open set. Then there is an $a>0$ such that, for all $h\in(0,a)$
  there is on open set $\kV_h\subseteq\w{\!T\-2\kM}$ such that
  \begin{enumerate}
    \item
      the tuple $(\kV_{\-3h},\partial^+_h,\partial^-_h)$ is a discrete
      tangent bundle of $\kM_0$;
    \item
      $\partial_h^\pm$ is a diffeomorphism from $\w{\kV_{\-3h}}$ to an
      open neighborhood of $\Delta(\kM_0\times\kM_0)$.
  \end{enumerate}
  Moreover, for all $v_m\in\w{\!T\-4_}m\kM_0$, there is a sufficiently small
  $h$ and $\w{\kV_{\-3h}}$ such that $v_m\in\w{\kV_{\-3h}}$.
\end{proposition}

Given a discretization of a tangent bundle, one can obtain, by
choosing $h$ small enough, a discrete tangent bundle which provides
discrete analogues of arbitrarily large tangent vectors. In physical
contexts, this implies that arbitrarily high velocities can be
accommodated in the discrete systems by using sufficiently small time
steps.

\begin{proof}[Proof of Proposition~\ref{prp:discretization-gives-discrete}]
In the generic manifold context the construct $m^+-m^-$ of
Equation~\eqref{eq:example-blowup} is unavailable, but it can be
replaced by the fibers of a tubular neighborhood of the diagonal
$\Delta(\kM\times\kM)$. The vector bundle $E\=\set{(v_m,-v_m)}{v_m\in
\w{\!T\-2\kM}}$ is a normal bundle to the diagonal $\Delta(\kM\times\kM)$ of
$\kM\times\kM$, so there is a tubular neighborhood $\zeta\colon
W^E\subset E\rightarrow W^{\kM\times\kM}\subset\kM\times\kM$.

The diffeomorphism $\zeta$ may be chosen so that $\!T\zeta$ is the
identity on the zero section $0(E)$ with respect to the
horizontal-vertical decomposition i.e.\ for all $w\in\w{\!T\-4_}{0_{(m,m)}}E$,
\[[eq:z-hor-vert]
  \!T\zeta(w)=\hor w+\vrt w.
\]
Any $(v_m^+,v_m^-)\in\w{\!T\-4_}{(m,m)}(\kM\times\kM)$ may be decomposed as
\[
  (v_m^+,v_m^-)=\left(\frac12(v_m^++v_m^-),\frac12(v_m^++v_m^-)\right)+
    \left(\frac12(v_m^+-v_m^-),\frac12(v_m^--v_m^+)\right).
\]
If $\!T\zeta^{-1}(v_m^+,v_m^-)=w$ then this is the unique
decomposition of $\!T\zeta(w)$ according to the direct sum
$\w{\!T\-4_}{(m,m)}(\kM\times\kM)\oplus E_{(m,m)}$. By
Equation~\eqref{eq:z-hor-vert}, $\hor w+\vrt w$ is also this
decomposition of $\!T\zeta(w)$, so
\[[eq:vert-T-zeta]
  \vrt\!T\zeta^{-1}(v_m^+,v_m^-)=
    \left(\frac12(v_m^+-v_m^-),\frac12(v_m^--v_m^+)\right).
\]

Consider the map $\varphi\colon\set{(h,v_m)}{\partial^\pm_h(v_m)\in
W^{\kM\times\kM}}\rightarrow\RR\times E$ by
\[
  \varphi(h,v_m)\=
  \begin{cases}
    \displaystyle\biggl(h,\frac1h\zeta^{-1}\,\partial^\pm_h(v_m)\biggr),
      &h>0,
    \\[10pt]
    \displaystyle\biggl(0,\left(\frac{v_m}2,-\frac{v_m}2\right)\biggr),&h=0.
  \end{cases}
\]
Using Equations~\eqref{101}, and~\eqref{eq:vert-T-zeta},
\[
  \vrt\frac d{dh}\biggr|_{h=0}\zeta^{-1}\,\partial^\pm_h(v_m)
    =\left(\frac{v_m}2,-\frac{v_m}2\right).
\]
so $\varphi$ is smooth
by~\cite{CuellC-PatrickGW-2007-1},~Proposition~1. $\varphi$ is a local
diffeomorphism at any $(0,v_m)$ since the derivative of $\varphi$ is
nonsingular there, and $\varphi$ is a diffeomorphism from
$\sset{0}\times\w{\!T\-2\kM}$ to $\sset{0}\times E$, so $\varphi$ is a
diffeomorphism from some open neighborhood of $\sset{0}\times\w{\!T\-2\kM}$
to some open neighborhood of $\sset{0}\times E$
(\cite{CuellC-PatrickGW-2007-1}, Theorem~1). The domain of the map
\[
  \tilde\varphi\bigl(h,(v_m,-v_m)\bigr)\=\bigl(h,\zeta(hv_m,-hv_m)\bigr)
\]
includes $\sset{0}\times E$, $\tilde\varphi$ is a diffeomorphism
except at $h=0$, and
\[
  \bigl(h,\partial^\pm(v_m)\bigr)=\tilde\varphi\,\varphi(h,v_m).
\]
Thus there are open neighborhoods $U\supseteq\sset{0}\times
\w{\!T\-2\kM}$ and $W\supseteq\sset{0}\times\Delta(\kM\times\kM)$
such that $(h,v_m)\mapsto\bigl(h,\partial_h^\pm(v_m)\bigr)$ is a
diffeomorphism from $U\setminus\bigl(\sset{0}\times
\w{\!T\-2\kM}\bigr)$ to $W\setminus\bigl(\sset{0}\times(\kM\times
\kM)\bigr)$.

Given a relatively compact open $\kM_0\subseteq\kM$, choose $a>0$ such
that $[0,a)\times\Delta(\kM_0\times\kM_0)\subseteq W$. Assume $0<h<a$
and define
\[
  \w{\kV_{\-3h}}\=\set{v_m}{(h,v_m)\in U}\cap(\partial^+_h)^{-1}(\kM_0)
    \cap(\partial^-_h)^{-1}(\kM_0).
\]
$\partial_h^\pm$ is a diffeomorphism on $\w{\kV_{\-3h}}$ since
$(h,v_m)\mapsto\bigl(h,\partial_h^\pm(v_m)\bigr)$ is a diffeomorphism.
Also, $\bigl(h,(m_0,m_0)\bigr)\in W$ for any $m_0\in\kM_0$, so
$\delta^+_h(v_m)=m_0$ and $\delta^-_h(v_m)=m_0$ where $v_m$ is defined
by $\tilde\varphi(h,v_m)=\bigl(h,(m_0,m_0)\bigr)$, hence
$\partial^+_h$ and $\partial^-_h$ are onto $\kM_0$. By continuity,
given any $v_m\in\w{\!T\-2\kM}_0$, $h$ can be chosen so small that
$\partial^+_h(v_m)\in\kM_0$ and $\partial^-_h(v_m)\in\kM_0$, and so
small that $(h,v_m)\in U$. Thus $h$ can be chosen so small that
$v_m\in\w{\kV_{\-3h}}$.\elsqed
\end{proof}

As an aside, we get the following Corollary, which, given a second
order vector field, is obtained by applying the proof of
Proposition~\ref{prp:discretization-gives-discrete} (particularly the
construction of $\tilde\varphi$) to the $X$~discretization with
bias~$\alpha^-(h)=0,\alpha^+(h)=h$.

\begin{corollary}\label{cor:Delta-second-order-ode}
  Let $X$ be a $C^k$ second order vector field on $\w{\!T\-2\kM}$,
  $k\ge2$. Then there are open neighborhoods $U\supseteq\sset{0}\times
  \w{\!T\-2\kM}$ and $W\supseteq\sset{0}\times\Delta(\kM\times\kM)$
  such that
  $(t,v)\mapsto\bigl(t,\w{\tau\-2_\kM}\+2\w{F^X_{\-6t}}(v),\w{\tau\-2_\kM}(v)\bigr)$
  is a $C^{k-1}$ diffeomorphism from $U\setminus\sset{0}\times
  \w{\!T\-2\kM}$ to $V\setminus\sset{0}\times\Delta(\kM\times\kM)$.
\end{corollary}
\noindent
Corollary~\ref{cor:Delta-second-order-ode} is important when
constructing classical generating functions of type~1 for Lagrangian
systems $L\colon\kQ\to\RR$. These are functions on $\kQ\times\kQ$
which are defined as integrals of the Lagrangian~$L$ over solutions
with specified endpoints i.e.\ the classical action as a function of
endpoints. To well define the generating function using the flow of
the Lagrangian vector field, one should construct the map
$\Delta_t(q_2,q_1)$ from $\kQ\times\kQ$ to $\!T\kQ$ that returns the
initial velocity at $q_1$ that evolves to $q_2$ over time
interval~$t$. The generating function is then
\[
  \w{S\-5_t}(q_2,q_1)\=\int_0^t\w{F^{X_E}_{\-6s}}\bigl(\Delta_t(q_2,q_1)\bigr)\,ds
\]
where $X_E$ is the Euler-Lagrange vector field and $F^{X_E}$ is its
flow. The map $\Delta(q_2,q_1)$ cannot be straightforwardly
constructed using the implicit function theorem (as is attempted for
example in~\cite{AbrahamR-MarsdenJE-RatiuTS-1988-1}) because of a
singularity at $t=0$: infinite velocity is required to traverse from
$q_1$ to $q_2$ in zero time. But the map $\Delta(q_2,q_1)$ is easily
extracted from Corollary~\ref{cor:Delta-second-order-ode}. The MV
discrete `exact' Lagrangian~\cite{MarsdenJE-WestM-2001-1} is the
same as the the type~1 generating function and both suffer the same
singularity.

In the context of $\kQ\=\RR^N$ with $\psi(h,t,v_x)\=x+tv$ and bias
$\alpha^-(h)\=0,\alpha^+(h)\=h$, one has
\[
  (\partial_h^\pm)^{-1}(m^-,m^+)=\bigl(m^-,(m^+-m^-)/h\bigr).
\]
Thus the inverse of $\partial_h^\pm$ may be given the interpretation
of a difference quotient. This is used in
Definition~\ref{def:discrete-derivative} to define the discrete
derivative of a sequence in $\kM$. That is important because it gives
the definition of a discrete first order sequence in $\w{\!T\-2\kM}$,
which is crucial to the discrete variation principle \LdApd\ of
Section~\ref{sec:discrete-LdA}.

\begin{definition}\mbox{}\label{def:discrete-derivative}
  \begin{enumerate}
    \item
      If $m_k$ is a sequence in $\kM$, then a \defemph{discrete
      derivative of $m_k$} is a sequence $m_k^\prime\in\kV$ such that
      $\partial^\pm(m^\prime_k)=(m_{k+1},m_k)$.
    \item
      A sequence $v_k\in\kV$ is called \defemph{first order} if
      $v_k=m_k^\prime$ for some sequence $m_k\in\kM$.
   \end{enumerate}
\end{definition}
\noindent A sequence $m_k$ is first order if and only if
\[[eq:first-order]
  \partial^+(v_k)=\partial^-(v_{k+1}),
\]
because the derivative of every sequence $m_k$ satisfies
Equation~\eqref{eq:first-order}, and every sequence $v_k$ satisfying
Equation~\eqref{eq:first-order} is the derivative of
\[
  m_0\=\partial^-(v_0),\quad m_1\=\partial^-(v_1),\ldots,
    m_k\=\partial^-(v_k),
    \ldots, m_{N-1}\=
    \partial^-(v_{N-1}),\quad m_N\=\partial^+(v_{N-1}).
\]
By Remark~\ref{rem:partial-pm-diffeo}, or
Proposition~\ref{prp:discretization-gives-discrete} if the maps
$\partial^+,\partial^-$ arise from a discretization, the discrete
derivative of $m_k$ is unique as long as the pairs $(m_{k+1},m_k)$ lie
sufficiently close to the diagonal and the sequence values $v_k$ are
restricted to be sufficiently near the zero section.

Let $(\kV,\partial^+,\partial^-)$ be a discrete tangent
bundle of $\kM$.
Define the \defemph{backward vertical bundle} by
\[
  \w{\vrt^+_v\-3}\kV\=\ker\w{\!T\-4_}v\partial^+,\qquad
  \w{\vrt^+\-3\kV}\=\ker\!T\partial^+,
\]
and the \defemph{forward vertical bundle} by
\[
  \w{\vrt^-_v\-3}\kV\=\ker\w{\!T\-4_}v\partial^-,\qquad
  \w{\vrt^-\-3\kV}\=\ker\!T\partial^-.
\]
The fibers of the forward [backward] vertical bundles are the tangent
spaces to the forward [backward] fibers of the discrete tangent
bundle. Item~\ref{enum:discrete-tangent-bundle-1} of
Definition~\ref{def:discrete-tangent-bundle} gives
$\!T\kV=\w{\vrt^+\-3\kV}\oplus\w{\vrt^-\-3\kV}$ so that every $\delta
v\in\w{\!T\-4_}v\kV$ decomposes uniquely as $\delta v=\delta
v^++\delta v^-$ where $\delta v^+\in{\vrt^-_v}\kV$ and $\delta
v^-\in\w{\vrt^+_v\-3}\kV$. The signs may appear notationally reversed
but they are mnemonic in the sense the one wants more often to apply
$\!T\partial^+$ not to elements of $\w{\vrt^-_v\-3}\kV$, which would
result in zero, but rather to elements of $\w{\vrt^+_v\-3}\kV$. The
convention $\delta v^+\in\w{\vrt^-_v\-3}\kV$ means that, usually,
`$+$' goes with `$+$' to make something nonzero, while zero results
when `$+$' goes with `$-$'.

\begin{figure}[ht]
  \begin{center}\input{f3_pstex_t.tex}\end{center}
\end{figure}

\begin{remark}
  Lagrangian discretizations are towards constructing discrete
  Lagrangian models which have a stature with continuous Lagrangian
  models. However, not every construct that is well defined in the
  context of continuous models is also well defined in the context of
  discrete models. In the diagram above, at left, any point of $\kM$
  near the area straddling $\partial^-(v)$ and $\partial^+(v)$ could
  be considered the base point of the discrete tangent vector $v$. The
  finite~---~as opposed to infinitesimal~---~nature of the discrete
  tangent vectors precludes an unambiguous association of
  configurations to elements of the discrete phase spaces. In the
  discrete context, the association of configurations to points of
  velocity phase space is artificial~---~like the invocation of a
  metric or connection where none is really natural. This, of course,
  is somewhat unintuitive after such concentration on the continuous
  systems. The reflex to associate a unique configuration to a
  velocity has to be unlearned.
\end{remark}

\begin{lemma}\label{lem:partials-are-isomorphisms}
  For all $v\in\kV$, $\w{\!T\-4_}v\partial^-$
  [resp.\ $\w{\!T\-4_}v\partial^+$] is a linear isomorphism from
  $\w{\vrt^+_v\-3}\kV$ [resp.\ $\w{\vrt^-_v\-3}\kV$] to
  $\w{\!T\-4_}{\partial^-(v)}\kM$
  [resp.\ $\w{\!T\-4_}{\partial^+(v)}\kM$].
\end{lemma}
\begin{proof}
If $\delta v\in\w{\vrt^+_v\-3}\kV$ and $\w{\!T\-4_}v\partial^-(\delta v)=0$ then
$\delta v\in\ker\!T\partial^+\cap\!T\partial^-$, so $\delta v=0$ by
Definition~\ref{def:discrete-tangent-bundle}. Thus $\w{\!T\-4_}v\partial^-$
has trivial kernel on $\w{\vrt^+_v\-3}\kV$ and the result follows because
$\dim\w{\vrt^+_v\-3}\kV=\dim\kM=\dim\w{\!T\-4_}m\kM$.\elsqed
\end{proof}
\noindent
In particular, if $\delta m\in\w{\!T\-4_m\kM}$ and $\partial^+(v)=m$
[resp.\ $\partial^-(v)=m$], then there is a unique $\delta
v\in\w{\vrt^-_v\-3}\kV$ [resp.\ $\delta v\in\w{\vrt^+_v\-3}\kV$] such that
$\!T\partial^+(\delta v)=\delta m$ [resp.\ $\!T\partial^-(\delta
v)=\delta m$]. If $v$ is understood, then we will write $\delta
v=\delta m^+$ [resp.\ $\delta v=\delta m^-$] i.e.\ $\delta m^+$
and $\delta m^-$ satisfy
\[
  \!T\partial^-(\delta m^+)=0,\qquad
  \!T\partial^+(\delta m^+)=\delta m,\qquad
  \!T\partial^-(\delta m^-)=\delta m,\qquad
  \!T\partial^+(\delta m^-)=0.
\]
Combining, if $v,\tilde v\in\kV$ are such that
$\partial^+(v)=m=\partial^-(\tilde v)$, then there are unique vectors
$\delta v\in\w{\vrt^-_v\-3}\kV$ and
$\delta\tilde v\in\w{\vrt^+_{\tilde v}\-3}\kV$ such that
\[
  \!T\partial^+(\delta v)=\delta m=\!T\partial^-(\delta\tilde v).
\]
This provides a linear isomorphism $\tau_{\tilde
  v,v}\colon\w{\vrt^-_v\-3}\kV\to\w{\vrt^+_{\tilde v}\-3}\kV$ as
follows: if $\delta v\in\w{\vrt^-_v\-3}\kV$, then $\tau_{\tilde
  v,v}(\delta v)\=\delta\tilde v$ is the unique vector in
$\w{\vrt^+_{\tilde v}\-3}\kV$ such that $\w{\!T\-4_}{\tilde
  v}\partial^-(\delta\tilde v)=\w{\!T\-4_}v\partial^+(\delta v)$.

Suppose that $(\kV,\partial^+,\partial^-)$ is a discrete tangent
bundle of $\kM$ and $\theta$ is a smooth one form on $\kV$. Let
$v\in\kV$, and set $m\=\partial^-(v)$ and $\tilde
m\=\partial^+(v)$. We will have use of two bilinear forms on
$\w{\!T\-4_}m\kM\times\w{\!T\-4_{\tilde m}\kM}$, denoted
$\!d^{\mp}\theta(v)$ and $\!d^{\pm}\theta(v)$, and defined as
follows. Given $\delta m\in\w{\!T\-4_}m\kM$ and $\delta\tilde m\in
\w{\!T\-4_{\tilde m}\kM}$, choose (local) vector fields $X$ and
$\tilde X$ with values in $\w{\vrt^+\-3\kV}$ and $\w{\vrt^-\-3\kV}$
respectively, such that
\[
  \!T\partial^-\bigl(X(m)\bigr)=\delta m,\qquad
  \!T\partial^-\,X(x)=\!T\partial^-\,X(y)\quad\mbox{if}\quad
    \partial^-(x)=\partial^-(y),
\]
and also this with $\delta\tilde m$, $\tilde X$, and `$+$' instead of
$\delta m$, $X$, and `$-$', respectively. Such vector fields $X$ and
$\tilde X$ commute, because
\[
  \!T\partial^-\,[\tilde X,X](m)
  =\frac d{dt}\biggr|_{t=0}\!T\partial^-\,\Bigl(\w{F^{\tilde X}_{\-6t}}\Bigr)^*X(m)
  =\frac d{dt}\biggr|_{t=0}\!T\partial^-\,X\Bigl(\w{F^{\tilde X}_{\-6t}}(m)\Bigr)=0.
\]
Define
\[[eq:def-dpm-dmp]
  \w{\!d^{\mp}\-3\theta}(v)(\delta m,\delta\tilde m)\=
    \bigl(X(\!\iota_{\tilde X}\theta)\bigr)(v),\qquad
  \!d^{\pm}\theta(v)(\delta m,\delta\tilde m)\=
    \bigl(\tilde X(\!\iota_X\theta)\bigr)(v).
\]
This well defines $\w{\!d^{\mp}\-3\theta}(v)$ : if $X^\prime$ and $\tilde
X^\prime$ are other choices of such vector fields, then, using the
identity
$\!\iota_{[V,W]}\alpha=\!L_V\!\iota_W\alpha-\!\iota_W\!L_V\alpha$,
\[
  \bigl(X(\!\iota_{\tilde X}\theta)\bigr)(v)
    =\!L_{X^\prime}\!\iota_{\tilde X}\theta(v)
    =\!\iota_{\tilde X}\!L_{X^\prime}\theta(v)
    =\!\iota_{\tilde X^\prime}\!L_{X^\prime}\theta(v)
    =X^\prime(\!\iota_{\tilde X^\prime}\theta)(v),
\]
and similarly $\!d^{\pm}\theta(v)$ is well defined. Also, note that
\[
  \!d\theta\bigl(X(v),\tilde X(v)\bigr)
  =X\bigl(\!\iota_{\tilde X}\theta\bigr)(v)
    -\tilde X\bigl(\!\iota_X\theta\bigr)(v)-\theta([X,\tilde X])(v)
  =\w{\!d^{\mp}\-3\theta}(v)(\delta m,\delta\tilde m)
    -\!d^{\pm}\theta(v)(\delta m,\delta\tilde m),
\]
so $\w{\!d^{\mp}\-3\theta}=\!d^{\pm}\theta$ if and only if $\theta$ is
closed.

%
\section{Lagrange--d'Alembert principle}
\label{sec:LdA-principle}
%
%
%
\subsection{Continuous Lagrange--d'Alembert principle}
\label{sec:LdA-continuous}
%
%
\noindent Let $\kQ$ be a manifold of system configurations, and $\kD\subseteq
\!T\kQ$ be a submanifold such that $\w{\tau\-2_\kQ}|\kD$ is a submersion. To
a given Lagrangian $L\colon\!T\kQ\to\RR$, the corresponding
\defemph{action functional} assigns to curves $q(t)\in\kQ$ the number
\[[eq:continuous-q-action]
  S\=\int_a^bL\bigl(q^\prime(t)\bigr)\,dt.
\]
The \defemph{variational derivative} of $S$ is
\[
  \!dS\bigl(q(t)\bigr)\,\delta q(t)\=
    \frac d{d\epsilon}\biggr|_{\epsilon=0}
    S\bigl(q_\epsilon(t)\bigr)=0,
\]
where $\delta q(t)$ is a curve in $\!T\kQ$, and $q_\epsilon(t)$ satisfies
\[
  q_\epsilon(t)\biggr|_{\epsilon=0}=q(t),\qquad
  \frac\partial{\partial\epsilon}\biggr|_{\epsilon=0}q_\epsilon(t)=\delta q(t).
\]
Let $\kE$ be a subbundle of the pull-back bundle
$(\w{\tau\-2_\kQ}|\kD)^*(\!T\kQ)$ i.e.\ a smooth assignment of subspaces of
$\w{\!T\-4_}q\kQ$ to each $v_q\in\kD$. By definition, the curve $q(t)$ is an
\defemph{evolution} if it satisfies $\mbox{\LdA}(L,\kD,\kE)$, defined
as:
\medskip
\[
  \left\{\begin{array}{l} \mbox{LdA-1 (given constraint):\quad
    $q^\prime(t)\in\kD$.}
    \\[5pt]
    \mbox{LdA-2 (criticality):\quad $\!dS\bigl(q(t)\bigr)\,\delta q(t)=0$ for
      all $\delta q$ which satisfy}
    \\[3pt]
    \qquad\mbox{LdA-2a (given constraint forces):\quad
      $\delta q(t)\in\kE_{q^\prime(t)}$;}
    \\[3pt]
    \qquad\mbox{LdA-2b (fixed boundary):\quad $\delta q(a)=0$ and
      $\delta q(b)=0$.}
\end{array}\right.
\]

\noindent This is the general version, where the annihilator of
$\kE_{v_q}$ is the vector space of the constraint forces at state
$v_q$, and where $\kD$ is a nonlinear constraint on velocities.

Since $\w{\tau\-2_\kQ}$ is assumed to be a submersion on $\kD$,
$\ker\bigl(\!T(\w{\tau\-2_\kQ}|\kD)\bigr)$ is a subbundle of $\vrt\!T\kD$
with fiber dimension $\dim\kD-\dim\kQ$, and
\[[eq:def-dot-D]
  \dot\kD_{v_q}\=\set{w_q\in\!T\kQ}{\w{\vrt_{v_q}\-3}w_q\in\!T\kD},
\]
is a subbundle of $(\w{\tau\-2_\kQ}|\kD)^*(\!T\kQ)$ with the same fiber
dimension. One possibility for $\kE$ is \defemph{Chetaev's rule}
$\kE\=\dot\kD$, but other choices may be appropriate, as discussed for
example in~\cite{MarleCM-1998-1}. Chetaev's rule specializes to the
usual case of \defemph{linear constraints} if $\kD$ is a distribution
on $\!T\kQ$, because then $\dot{\kD}_{v_q}=\kD_q$.

The Lagrange--d'Alembert principle above is written for curves
$q(t)\in\kQ$. We now transform it a variational principle for curves
$v(t)\in\!T\kQ$, by placing $q(t)$ and $v(t)$ in one-to-one
correspondence using $q(t)=\w{\tau\-2_\kQ}\,v(t)$ and
$v(t)=q^\prime(t)$. The transformed variational principle has the
additional constraint (the first order constraint)
$v(t)=\bigl(\w{\tau\-2_\kQ}\,v(t)\bigr)^\prime$ on curves
$v(t)\in\!T\kQ$. Substituting $q^\prime(t)=v(t)$ into
Equation~\eqref{eq:continuous-q-action} transforms the action to
\[
  S\=\int_a^bL\bigl(v(t)\bigr)\,dt,
\]
and we extend $S$ to all curves $v(t)$ by this same formula.
The \defemph{variational derivative} of $S$ is
\[
  \!dS\bigl(v(t)\bigr)\,\delta v(t)\=
    \frac\partial{\partial\epsilon}\biggr|_{\epsilon=0}
    S\bigl(v_\epsilon(t)\bigr),
\]
where $\delta v(t)$ is a curve in $\!T\!T\kQ$, and $v_\epsilon(t)$ satisfies
\[
  v_\epsilon(t)\biggr|_{\epsilon=0}=v(t),\qquad
  \frac\partial{\partial\epsilon}\biggr|_{\epsilon=0}
    v_\epsilon(t)=\delta v(t).
\]

The variation $v_\epsilon(t)$ implies a variation
$q_\epsilon(t)=\w{\tau\-2_\kQ}\,v_\epsilon(t)$; differentiating this
in~$\epsilon$ gives $\delta q(t)\=\!T\w{\tau\-2_\kQ}\,\delta v(t)$,
and thus the constraint $\delta q(t)\in\kE_{v(t)}$. The first order
constraint gives
$v_\epsilon(t)=\bigl(\w{\tau\-2_\kQ}\,v_\epsilon(t)\bigr)^\prime$ and
differentiating in $\epsilon$ gives the constraint on $\delta v(t)$
corresponding to the first order constraint:
\[
  \delta v(t)=\frac\partial{\partial\epsilon}
    \biggr|_{\epsilon=0}v_\epsilon(t)
  =\frac\partial{\partial\epsilon}\biggr|_{\epsilon=0}
    \,\frac\partial{\partial t}\w{\tau\-2_\kQ}\,v_\epsilon(t)
  =s_\kQ\,\frac d{dt}\!T\w{\tau\-2_\kQ}\,\delta v(t)=s_\kQ\,\delta q(t)^\prime,
\]
where $s_\kQ$ is the canonical involution on $\!T\!T\kQ$. Thus, the
Lagrange--d'Alembert principle transforms to $\mbox{\LdAp}(L,\kD,\kE)$:
\medskip
\[
  \left\{\begin{array}{l}
    \mbox{\LdAp-1 (constraints):}
    \\[3pt]
    \qquad\mbox{\LdAp-1a (given constraint):\quad
      $\displaystyle v(t)\in\kD$;}
    \\[3pt]
    \qquad\mbox{\LdAp-1b (first order constraint):\quad
      $\displaystyle v(t)=\bigl(\w{\tau\-2_\kQ}\,v(t)\bigr)^\prime$.}
    \\[5pt]
    \mbox{\LdAp-2 (criticality):\quad
      $\!dS\bigl(v(t)\bigr)\,\delta v(t)=0$ for all $\delta v$ which satisfy}
    \\[3pt]
    \qquad\mbox{\LdAp-2a (given constraint forces):\quad
      $\delta q(t)\in\kE_{v(t)}$, where
      $\delta q\equiv\!T\w{\tau\-2_\kQ}\,\delta v(t)$;}
    \\[3pt]
    \qquad\mbox{\LdAp-2b (fixed boundary):\quad
      $\delta q(a)=0$ and $\delta q(b)=0$;}
    \\[3pt]
    \qquad\mbox{\LdAp-2c (first order constraint):\quad
      $\delta v=s_\kQ\,\delta q^\prime$.}
  \end{array}\right.
\]
\noindent This transformed principle \LdAp\ has some technical
advantages and is better suited to construct discretizations where the
discrete states are elements of $\!T\kQ$. It has been used
in~\cite{PatrickGW-2006-1}, where the higher dimension of $\!T\kQ$ as
opposed to $\kQ$ provides some required freedom in a desingularization
of continuous Lagrangian systems at time interval zero.

\begin{remark}\label{rem:HP1}
Implementing the first order constraint as a
Lagrange multiplier gives
\[[eq:HP]
  S\=\int_a^bL\bigl(v(t)\bigr)+\bigl\langle p(t),q^\prime(t)-v(t)
    \bigr\rangle\,dt
\]
where naturally $p(t)\in\w{\!T^*\-4\kQ}$. This is the
\defemph{Hamilton-Pontryagin principle}
\cite{YoshimuraH-MarsdenJE-2006-1,YoshimuraH-MarsdenJE-2006-2}. The HP
principle variationally identifies the Legendre transform as the
Lagrange multiplier of the first order constraint i.e.\ it implies the
constraint $p=\w{\!F\-2L}$. See Section~\ref{sec:discrete-HP} for a
few more comments on discretizations of the HP principle.
\end{remark}

%
\subsection{Discrete Lagrange--d'Alembert principle}
\label{sec:discrete-LdA}
%
%
\noindent We will develop a variational principle for sequences of
points in a discrete tangent bundle $\kV$, analogously with the
continuous \LdAp\ principle of Section~\ref{sec:LdA-continuous}, which
is a variational principle for curves with values in $\!T\kQ$.

Let $\kQ$ be a configuration manifold and
$(\kV,\partial^+,\partial^-)$ a discrete tangent bundle on $\kQ$. The
underlying structure for the discrete variational principle, which we
call the \defemph{Lagrange--d'Alembert principle} is as follows:
\begin{itemize}
  \item
    Given a \defemph{(discrete) Lagrangian $L_d\colon\kV\to\RR$}, the
    corresponding \defemph{discrete action functional} assigns to
    sequences $v_d(k)\in\kV$, $k=1,\ldots,N$, the number
    \[
      \w{S_{\-6d,N}}(v_d)\=\sum_{k=1}^NL_d\bigl(v_d(k)\bigr).
    \]
    We have reserved the subscript $d$ to distinguish the discrete and
    continuous contexts. The derivative of $\w{S\-5_d}$ is
    \[
      \!d\w{S_{\-6d,N}}(v_d)\,\delta v_d\=\frac d{d\epsilon}\biggr|_{\epsilon=0}
        \w{S_{\-6d,N}}(v_{d,\epsilon})
    \]
    where $\delta v_d\in\!T\kV[1,N]$, and $v_{d,\epsilon}\in\kV[1,N]$
    satisfies
    \[
      v_{d,0}(k)=v_d(k),\qquad
      \frac\partial{\partial\epsilon}\biggr|_{\epsilon=0}
        v_{d,\epsilon}(k)=\delta v_d(k).
    \]
  \item
    The discrete velocity constraint is provided by a submanifold
    $\kD_d$ of $\kV$, such that $\partial^+|\kD_d$ and
    $\partial^-|\kD_d$ are submersions.
  \item
    In the continuous context, the constraint forces are determined by
    an association $\kE$ of subspaces of $\!T\kQ$ to velocities in the
    continuous constraint $\kD$. For the discrete context we assume a
    subbundle~$\kE_d$ of the pullback bundle
    $(\partial^+|\kD_d)^*(\!T\kQ)$ i.e.\ $\kE_d$ is a smooth
    assignment of subspaces of $\w{\!T\-4_}{\partial^+(v)}\kQ$ to
    $v\in\kD_d$.
  \item
    Higher order of accuracy of discretizations of (continuous)
    nonholonomically constrained Lagrangian systems require discrete
    analogues that do not fit the pattern just described. To
    accommodate this, we generalize and replace $\!dL_d$ with a one
    form $\w{\sigma\-6_d}$ on $\!T\kV$ and replace the derivative of the
    action $\!d\w{S\-6_N}$ with $\Sigma_{d,N}$ where
    \[
      \Sigma_{d,N}(v_d)\,\delta v_d\=\sum_{k=1}^N\w{\sigma\-6_d}
        \bigl(v_d(k)\bigr)\,\delta v_d(k).
    \]
    The net effect is that the discrete analogue $\w{\sigma\-6_d}$ of the
    derivative of the Lagrangian is not necessarily closed, and it
    contributes to the discrete analogue of the derivative of the
    action, which also is not necessarily closed.
\end{itemize}
A \defemph{discrete constrained Lagrangian system (DCLS)} is a tuple
$(\kV,\w{\sigma\-6_d},\kD_d,\kE_d)$ as above, where
$\kV\=(\kV,\partial^+,\partial^-)$ is a discrete tangent bundle. If
the constraint is absent, then the tuple $(\kV,\w{\sigma\-6_d})$ is simply a
\defemph{discrete Lagrangian system (DLS)}. By definition, a sequence
$v(k)$ is an \defemph{evolution} if it satisfies
\[
  \left\{\begin{array}{l}
    \mbox{\LdApd-1 (constraints):}
    \\[3pt]
    \qquad\mbox{\LdApd-1a (given constraint):\quad
      $\displaystyle v(k)\in\kD_d$;}
    \\[3pt]
    \qquad\mbox{\LdApd-1b (first order constraint):\quad
      $\partial^+\bigl(v_d(k)\bigr)=\partial^-\bigl(v_d(k+1)\bigr)$.}
    \\[5pt]
    \mbox{\LdApd-2 (criticality):\quad
    $\Sigma_{d,N}\bigl(v_d(k)\bigr)\,\delta v(k)=0$ for
      all $\delta v(k)$ which satisfy}
    \\[3pt]
    \qquad\mbox{\LdApd-2a (given constraint forces):\quad
      $\!T\partial^+\bigl(\delta v_d(k)\bigr)\in(\kE_{d})_{v_d(k)}$;}
    \\[3pt]
    \qquad\mbox{\LdApd-2b (fixed boundary):\quad
      $\!T\partial^-\bigl(\delta v(1)\bigr)=0$ and
      $\!T\partial^+\bigl(\delta v(N)\bigr)=0$;}
    \\[3pt]
    \qquad\mbox{\LdApd-2c (first order constraint):\quad
      $\!T\partial^+\bigl(\delta v_d(k)\bigr)=\!T\partial^-\bigl(\delta
      v_d(k+1)\bigr)$.}
  \end{array}\right.
\]
\smallskip
\noindent
This is the \defemph{discrete Lagrange--d'Alembert principle}
$\mbox{\LdApd}(\kV,\w{\sigma\-6_d},\kD_d,\kE_d)$.

\begin{definition}
  \[
    &\kC_{d,N}\=\set{w\in\kV[1,N]}{
        \partial^+\bigl(w(k)\bigr)=\partial^-\bigl(w(k+1)\bigr)};
    \\
    &\kN_{d,N}\=\set{w\in\kC_{d,N}}{w(k)\in\kD_d};
    \\
    &\kW_{d,N}\=\set{\delta w\in\w{\!T\-4_}w\kC_{d,N}}{
      \!T\partial^-\bigl(\delta w(1)\bigr)=0,\;
      \!T\partial^+\bigl(\delta w(N)\bigr)=0,\;
      \!T\partial^+\bigl(\delta w(k)\bigr)\in(\kE_d)_w}.
  \]
\end{definition}
\noindent Altogether, a sequence $v_d$ satisfies \LdApd\ if and only if
\begin{enumerate}
  \item
    $v_d$ lies in the manifold $\kC_{d,N}$; and
  \item
    $v_d$ satisfies the constraint $v_d\in\kN_{d,N}$; and
  \item
    $v_d$ is \defemph{critical}, meaning
    $\Sigma_{d,N}\bigl(v_d\bigr)\,\delta v_d=0$ for all $\delta
    v_d\in\kW_{d,N}$.
\end{enumerate}
The restriction to the first order submanifold $\kC_{d,N}$ is
implemented first. In part this is because the distribution
$\kW_{d,N}$ has no natural extension away from $\kC_{d,N}$, since it
is the \emph{common value} of the backward and forward projected
variations that have to be in~$\kE$.

\begin{remark}\label{rem:skew-ordinary-definition}
\LdApd~is potentially a skew critical problem, in that one seeks points
in a constraint~---~$\kN_{d,N}$~---~where a one
form~---~$\Sigma_{d,N}$~---~annihilates a
distribution~---~$\kW_{d,N}$~---~but that distribution is not
necessarily the tangent bundle of the constraint.
See~\cite{CuellC-PatrickGW-2007-1} and
Section~\ref{sec:LdA-skew-critical-problem}.
\end{remark}

As shown in Theorem~\ref{thm:reduce-to-N=2} below, the discrete
Lagrange--d'Alembert principle for sequences of arbitrary length is
equivalent to the same principle for consecutive pairs of the
sequence. This is critical to the construction of integrators, because
it reduces an optimization on length~$N$ sequences to an iteration on
length~$N=2$ sequences.
Thus, the $N=2$ case occurs often, and it is
helpful to abbreviate its notations.
\begin{definition}
\[
  &\Sigma_d(v,\tilde v)\=\w{\sigma\-6_d}(v)+\w{\sigma\-6_d}(\tilde v),
  \\[2pt]
  &\kC_d\=\set{(v,\tilde v)\in\kV\times\kV}{\partial^+(v)
    =\partial^-(\tilde v)},
  \\[2pt]
  &\kN_d\=\set{(v,\tilde v)\in\kC_d}{v,\tilde v\in\kD_d},
  \\[2pt]
  &\kW_d\=\set{(\delta v,\delta\tilde v)\in\!T\kC_d}{
    \!T\partial^-(\delta v)=0,\;
    \!T\partial^+(\delta\tilde v)=0,\;
    \!T\partial^+(\delta v)\in(\kE_d)_v}.
\]
i.e.\ the atomic $N=2$ case is abbreviated by by dropping the $N$
subscript. $(v,\tilde v)$ is a \defemph{solution pair} if it
satisfies~\LdAp for $N=2$.
\end{definition}

\begin{theorem}\label{thm:reduce-to-N=2}
  Let $(\kV,\w{\sigma\-6_d},\kD_d,\kE_d)$ be a DCLS. Then a sequence
  $v_d(k)$ is a discrete evolution if and only if each pair $(v,\tilde
  v)=\bigl(v_d(k),v_d(k+1)\bigr)$ is a discrete evolution, $1\le k\le
  N-1$.
\end{theorem}

\begin{proof}
Obviously, $v_d\in\kN_{d,N}$ if and only if every pair
$\bigr(v_d(k),v_d(k+1)\bigl)\in\kN_d$. So it is only necessary to show
that, for $v_d\in\kN_{d,N}$, $\Sigma_{d,N}(v_d)$ annihilates every
$\delta v_d\in(\kW_{d,N})_{v_d}$ if and only if
$\Sigma_d\bigl(v(k),v(k+1)\bigr)$ annihilates every $(\delta
v,\delta\tilde v)\in(\kW_d)_{(v(k),v(k+1))}$, $1\le k\le N-1$.

A sequence $\delta v_d\in\kW_{d,N}$ has $\delta v_d(1)^-=0$ and
$\delta v_d(N)^+=0$ by definition, and so is the sum of the sequences
which are the rows of the following array:
\[
  \begin{array}{ccccccc}
    \delta v_d(1)^+&\delta v_d(2)^-&0&0&\cdots&0&0
    \\
    0&\delta v_d(2)^+&\delta v_d(3)^-&0&\cdots&0&0
    \\
    0&0&\delta v_d(3)^+&\delta v_d(4)^-&\cdots&0&0
    \\
    \vdots&\vdots&\vdots&\vdots&&\vdots&\vdots
    \\
    0&0&0&0&\cdots&\delta v_d(N-1)^+&\delta v_d(N)^-
  \end{array}
\]
As is easily verified, this corresponds to the direct sum
decomposition
\[[eq:WdN-direct-sum-Wd]
  \kW_{d,N}=\bigoplus_{k=1}^{N-1}(\kW_d)_{(v(k),v(k+1))}
\]
where elements of the subspaces $(\kW_d)_{(v(k),v(k+1))}$ are
understood to be, after appropriate padding with zeros, sequences of
length $N$. Thus $\Sigma_{d,N}$ vanishes on $\kW_{d,N}$ if and only if
$\Sigma_{d,N}$ vanishes on each factor $(\kW_d)_{(v(k),v(k+1))}$.
But, for $\delta v_d\in\kW_{d,N}$,
\[
  \Sigma_{d,N}(v_d)\,\delta v_d
    &=\sum_{k=1}^N\w{\sigma\-6_d}\bigl(v_d(k)\bigr)\,\delta v_d(k)
  \\
    &=\sum_{k=1}^N\w{\sigma\-6_d}\bigl(v_d(k)\bigr)\bigl(\delta v_d(k)^-
      +\delta v_d(k)^+\bigr)
  \\
    &=\w{\sigma\-6_d}\bigl(v_d(1)\bigr)\,\delta v_d(1)^-+\w{\sigma\-6_d}
      \bigl(v_d(N)\bigr)\,\delta v_d(N)^+
    +\sum_{k=1}^{N-1}\Bigl(\w{\sigma\-6_d}\bigl(v_d(k+1)\bigr)
      \,\delta v_d(k+1)^-+\w{\sigma\-6_d}\bigl(v_d(k)\bigr)\,
      \delta v_d(k)^+\Bigr)
  \\
    &=\sum_{k=1}^N\Sigma_d\bigl(v_d(k),v_d(k+1)\bigr)
      \bigl(\delta v_d(k)^+,\delta v_d(k+1)^-\bigr),
\]
i.e.\ with respect to the decomposition~\eqref{eq:WdN-direct-sum-Wd},
$\Sigma_{d,N}=\bigoplus_{k=1}^{N-1}\Sigma_d$.\elsqed
\end{proof}

\begin{remark}\label{rm:localization}
Many fundamental physical systems have continuous variational
formulations with a fixed boundary constraint, and with action defined
as an integral of a local Lagrangian. The solutions of such
variational formulations have the essential property of
\defemph{localization}: restrictions of solutions are solutions. This
follows directly from the variational principle. Indeed, the action of
a solution is a sum of the action over the restriction of a solution
and the complement of that, and a fixed boundary variation of such a
restriction is a variation of the whole. So the restriction is
critical under such variations, because under them the whole is
critical and the action is constant on the complement of the
restriction. The proof of Theorem~\ref{thm:reduce-to-N=2}, which is
also purely variational, shows that the discrete skew critical
problem, for arbitrarily long sequences, is equivalent to successive
skew critical problems, for sequences of length 2. This is because the
discrete action is a sum over $\w{\sigma\-6_d}$, and because of the
fixed boundary constraint. Thus the discrete systems have
localization to the discretization scales for the same reasons that
the continuous systems have localization to arbitrary scales.
\end{remark}

%
\subsection{The discrete Hamilton-Pontryagin principle}
\label{sec:discrete-HP}
%
%
\noindent One approach to discretizations of Lagrangian systems is
through discretizations of the Hamilton-Pontryagin
principle~\eqref{eq:HP} as
in~\cite{BouRabeeN-MarsdenJE-2008-1,LallS-WestM-2006-1}. The
Hamilton-Pontryagin principle does not immediately discretize in the
formalism of this article, because
\begin{enumerate}
\item
  it requires the difference $q^\prime(t)-v(t)$, but discrete tangent
  bundles do not support linear operations; and
\item
  it requires $q(t)$ from $v(t)$, whereas there is no unique
  projection to configurations from discrete tangent bundles.
\end{enumerate}

To recover the HP principle in our context, one might posit additional
constructs sufficient to intrinsically write the principle itself. For
example, a discrete analogue of the difference $q^\prime(t)-v(t)$
could be constructed using an appropriate submersion
$\Delta\colon\kV\times\kV\to\!T\kQ$. We choose not to pursue this here,
but rather note that one can an apply Lagrange multipliers to the
variations, \emph{after} differentiating the action and \emph{after}
imposing the second order constraint in phase space. That is, the
discrete HP principle is obtained by removing~\LdApd-2c and
replacing~\LdApd-2 with
\[
  \Sigma_{d,N}\bigl(v_d(k)\bigr)\,\delta v(k)
    +\bigl\langle p_k,\!T\partial^-\bigl(\delta v_d(k+1)\bigr)
    -\!T\partial^+\bigl(\delta v_d(k)\bigr)\bigr\rangle=0.
\]
The difference is valid because it occurs in the single tangent fiber
of $\!T\kQ$ at
$\partial^+\bigl(v_d(k)\bigr)=\partial^-\bigl(v_d(k+1)\bigr)$.
Reverting to $N=2$ gives
\[
  \w{\sigma\-6_d}(v)\,\delta v+\w{\sigma\-6_d}(\tilde v)\,\delta\tilde v
    +\langle p,\!T\partial^-(\delta\tilde v)\rangle
    -\langle p,\!T\partial^+(\delta v)\rangle=0,
\]
or, putting separately $\delta v=\delta q^+,\delta\tilde v=0$ and then
$\delta v=0,\delta\tilde v=\delta q^-$,
\[
  \langle p,\delta q\rangle=\langle\w{\sigma\-6_d}(v),\delta q^+\rangle,\qquad
  \langle p,\delta q\rangle=-\langle\w{\sigma\-6_d}(\tilde v),\delta q^-\rangle.
\]
This identifies the discrete Legendre transforms
\[
  \langle\w{\!F^+\-5L}(v),\delta q\rangle\=\langle\w{\sigma\-6_d}(v),
    \delta q^+\rangle,\qquad
  \langle\w{\!F^-\-4L}(v),\delta q\rangle\=-\langle\w{\sigma\-6_d}(v),
    \delta q^-\rangle,
\]
and we have the commutative diagrams
$$
\includegraphics{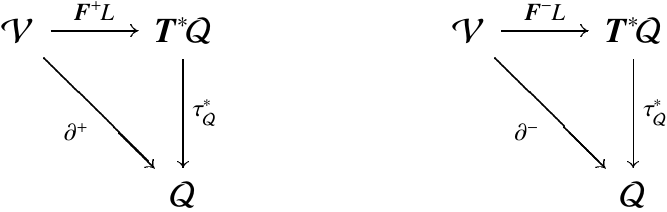}
$$

%
\section{Equations of motion}
%

%
\subsection{Continuous equations of motion}
\label{sec:continuous-equations-motion}
%
%
\noindent In the continuous context, localization as explained in
Remark~\ref{rm:localization} lends to the expectation of differential
equations of
motion~\cite{MarsdenJE-PatrickGW-ShkollerS-1998-1}. Defining the
\defemph{second order submanifold}
\[
  \ddot\kQ\=\set{q^{\prime\prime}(0)}
    {\mbox{$q(t)$ a $C^2$ curve in $\kQ$}},
\]
there is a unique section $\partial L$ of
$\hom(\!T\kQ,\w{\!T^*\-4\kQ})$ and a unique section $\delta
L\colon\ddot\kQ\rightarrow\w{\!T^*\-4\kQ}$ of the bundle
$(\w{\tau\-1_{\!T\kQ}}|\ddot\kQ)^*(\w{\!T^*\-4\kQ})$, such that
\[
  \!dS\bigl(q(t)\bigr)\,\delta q(t)
    =\int_a^b\delta L\bigl(
    q^{\prime\prime}(t)\bigr)\,\delta q(t)\,dt
    +\partial L(q^\prime(t)\bigr)\,\delta q\biggr|_a^b.
\]
Defining the one form $\theta_L(v_q)w_{v_q}\=\partial
L(v_q)\,\!T\w{\tau\-2_\kQ}(w_{v_q})$,
\[[eq:dS-continuous]
  \!dS\bigl(q(t)\bigr)\,\delta q(t)
    =\int_a^b\delta L\bigl(
    q^{\prime\prime}(t)\bigr)\,\delta q(t)\,dt
    +\theta_L\bigl(q^\prime(t)\bigr)\,\delta v\biggr|_a^b,
\]
where
\[
  \delta q(t)\=\frac\partial{\partial\epsilon}
    \biggr|_{\epsilon=0}q_\epsilon(t),\qquad
    \delta v(t)\=\frac\partial{\partial\epsilon}
    \biggr|_{\epsilon=0}\,\frac\partial{\partial t}
    \biggr|_{t=0}q_\epsilon(t).
\]
So the variational principle identifies $\delta L$ and $\theta_L$
directly, and a curve $q^i(t)$ is an evolution if and only it
satisfies
\[[eq:continuous-LdA-Q]
  q^\prime(t)\in\kD,\qquad\delta L\bigl(q^\prime(t)\bigr)
    =-\lambda(t)\in\ann\kE.
\]
These are the \defemph{(continuous) Lagrange--d'Alembert equations}
for curves in $\kQ$. In coordinates, $\ddot\kQ=\sset{(q^i,\dot
q^i,\dot q^i,\ddot q^i)}$, and
\[
  \delta L=\biggl(\frac{\partial L}{\partial q^i}-\frac d{dt}
    \frac{\partial L}{\partial\dot q^i}\biggr)\,dq^i
    =\biggl(\frac{\partial L}{\partial q^i}
    -\frac{\partial^2L}{\partial\dot q^i\partial q^j}\dot q^j
    -\frac{\partial^2L}{\partial\dot q^i\partial\dot q^j}\ddot q^j
    \biggr)\,dq^i,
\]
so that a curve $q^i(t)$ is an evolution if and only it satisfies
the familiar
\[
  \frac{dq^i}{dt}\in\kD,\qquad\frac{\partial L}{\partial q^i}-\frac d{dt}
    \frac{\partial L}{\partial\dot q^i}
    =-\lambda_i(t)\in\ann\kE.
\]
All this corresponds to the principle \LdA~i.e.\ for curves on $\kQ$.
To cast it to the form of \LdAp~i.e.\ for curves on $\!T\kQ$, one must
assume that $v(t)$ is first order, or else the integration-by-parts
inherent in Equation~\eqref{eq:dS-continuous} will fail. Under that
restriction the various formulae transform easily, and
Equations~\eqref{eq:continuous-LdA-Q} become
\[
  v(t)=\bigl(\w{\tau\-2_\kQ}\,v(t)\bigr)^\prime,\qquad
  v(t)\in\kD,\qquad\delta L\bigl(v(t)\bigr)=-\lambda(t)\in\ann\kE.
\]
These are the \defemph{(continuous) Lagrange--d'Alembert equations}
for curves in $\!T\kQ$.

The \defemph{first and second fiber
  derivatives}~\cite{AbrahamR-MarsdenJE-1978-1} of $L$ are the maps
$\w{\!F\-2L}\colon\!T\kQ\to\w{\!T^*\-4\kQ}$ and
$\w{\!F^2\-4L}\colon\!T\kQ\to\!T^2_0\kQ$ defined by
\[
  \w{\!F\-2L}(v_q)w_q\=\!D(L|\w{\!T\-4_}q\kQ)(v_q)w_q,\qquad
  \w{\!F^2\-4L}(v_q)(w_q,\tilde w_q)\=\!D^2(L|\w{\!T\-4_}q\kQ)(v_q)(w_q,
    \tilde w_q).
\]
A Lagrangian $L$ is called \defemph{$(\kD,\kE)$-regular} if, for all
$v_q\in\kD$, the condition: $w_q\in\dot{\kD}_{v_q}$ (recall
Equation~\eqref{eq:def-dot-D}) and $\w{\!F^2\-4L}(v_q)(w_q,\tilde
w_q)=0$ for all $\tilde w_q\in\kE_{v_q}$ implies $w_q=0$.  If $L$ is
$(\kD,\kE)$-regular, then the fiber dimension of $\dot\kD$ and the
fiber dimension of $\kE$ are necessarily equal i.e.\ regularity
implies
\[
  \dim\kD-\dim\kQ=\fdim\kE
\]
or equivalently
\[
  2\dim\kQ-\dim\kD=\dim\kQ-\fdim\kE.
\]
The number of constraints in \LdA~(or its equivalent~\LdAp) is the
codimension of $\kD$ in $\!T\kQ$, while the dimension of the space of
constraint forces is the fiber dimension of the annihilator subbundle
of $\kE$ in $(\w{\tau\-2_\kQ})^*(\!T\kQ)$ i.e.\ there are
\[
  2\dim\kQ-\dim\kD,\qquad\dim\kQ-\fdim\kE
\]
independent constraints and constraint forces, respectively. At the
outset of \LdA, $\kD$ and $\kE$ are hypothesized and independent,
and the number of independent constraints is unrelated to the number
of independent constraint forces. Given regularity, equality of these
is assured, and we set
\[
  r\=\dim\kD-\dim\kQ=\fdim\kE,
\]
and also  there is~\cite{PatrickGW-2007-1}
a unique second order vector field~$Y_{\delta L}$ on $\kD$ such that
$\delta L\circ Y_{\delta L}$ annihilates $\kE$. Existence and
uniqueness for \LdA\ on the phase space $\kD$ follows because its
evolutions are the integral curves of $Y_{\delta L}$.

%
\subsection{Discrete equations of motion}
\label{sec:discrete-equations-motion}
%
%
\noindent To develop discrete equations of motion, we make the
following definitions:
\begin{enumerate}
  \item
    $\ddot\kQ_d\=\set{(v,\tilde v)}{\partial^+(v)=\partial^-(\tilde
    v)}$, which is a submanifold of $\kV\times\kV$. This is set
    theoretically the same as $\kC_d$, however for $\ddot\kQ_d$ we
    consider $\kV\times\kV$ to be a discrete tangent bundle of $\kV$,
    whereas for $\kC_d$ we consider $\kV\times\kV$ to be the atomic
    two-point evolutions in $\kV$.
  \item
    $\delta\w{\sigma\-6_d}(v,\tilde v)\,\delta
    q\=\w{\sigma\-6_d}(\tilde v)\,\delta
    q^-+\w{\sigma\-6_d}(v)\,\delta q^+$, which is a section of the
    pullback bundle $\pi_d^*\,\!TQ$, where $\pi_d\colon\ddot\kQ\to Q$
    by $\pi_d(v,\tilde v)=\partial^+(v)=\partial^-(\tilde v)$.
  \item
    $\theta_{\w{\sigma\-6_d}}^-(v)\,\delta v\=-\w{\sigma\-6_d}(v)\,\delta v^-$,
    $\theta_{\w{\sigma\-6_d}}^+(v)\,\delta v\=\w{\sigma\-6_d}(v)\,\delta v^+$,
    which are both one forms on $\kV$.
  \item
    $\omega_{\w{\sigma\-6_d}}^-\=-\!d\theta_{\w{\sigma\-6_d}}^-$ and
    $\omega_{\w{\sigma\-6_d}}^+\=\!d\theta_{\w{\sigma\-6_d}}^+$.
\end{enumerate}

\begin{remark}\label{rem:one-omega}
  $\theta_{\w{\sigma\-6_d}}^+-\theta_{\w{\sigma\-6_d}}^-=\w{\sigma\-6_d}$ so
  $\omega_{\w{\sigma\-6_d}}^+=\omega_{\w{\sigma\-6_d}}^-$ if $\w{\sigma\-6_d}$
  is closed, and in this case we write $\omega_{\w{\sigma\-6_d}}$ for either.
\end{remark}

From the proof of Theorem~\ref{thm:reduce-to-N=2}, the discrete
analogue of Equation~\eqref{eq:dS-continuous} is
Equation~\eqref{eq:dS-discrete} below. This is a critical equation for
our development and so it is separated here as a theorem.
\begin{theorem}\label{th:DCLS-fundamental-decomposition}
  If $(\kV,\w{\sigma\-6_d},\kD_d,\kE_d)$ is  a DCLS then
  \[[eq:dS-discrete]
    \iota_{\kC_{d,N}}^*\Sigma_{d,N}(v_d)\,\delta v_d
      =\sum_{k=1}^{N-1}\delta\w{\sigma\-6_d}(v_k,v_{k+1})\,\delta q(k)
        +\theta_{\w{\sigma\-6_d}}^+(v_N)\,\delta v(N)
        -\theta_{\w{\sigma\-6_d}}^-(v_1)\,\delta v(1)
  \]
  where $\iota_{\kC_{d,N}}\-2\colon\kC_{d,N}\to\kV[1,N]$ is the
  inclusion and $\delta q\=\!T\partial^+\bigl(\delta
  v_d(k)\bigr)=\!T\partial^-\bigl(\delta v_d(k+1)\bigr)$.
\end{theorem}
\noindent
Thus, a sequence $v(k)$ is an evolution if and only if it consists of
pairs $(v,\tilde v)$ which satisfy
\[[eq:discrete-LDA]
  \partial^+(v)=\partial^-(\tilde v),\qquad
  v,\tilde v\in\kD_d,\qquad
  \delta\w{\sigma\-6_d}(v,\tilde v)\in\ann\kE_d.
\]
These are the \defemph{discrete Lagrange--d'Alembert equations}.

There is no general existence and uniqueness result for the nonlinear
algebraic Equations~\eqref{eq:discrete-LDA}. However, local existence
and uniqueness can be analyzed at the level of linearizations, using
the inverse function theorem: View~\eqref{eq:discrete-LDA} as
equations for $\tilde v\in\kD_d$ given fixed $v\in\kD_d$, and denote
$q\=\partial^+(v)$. Smoothly choose (local) vector fields $X_{\delta
  q}$ on $\kQ$, linearly parametrized by elements of $\delta
q\in(\kE_d)_v$, such that $X_{\delta q}(q)=\delta q$ i.e.\ the vector
fields $X_{\delta q}$ are extensions of $\delta q\in(\kE_d)_v$. Each
$X_{\delta q}$ lifts to a vector field $X_{\delta q}^-$ taking values
in $\w{\vrt^+\-3\kV}$ and such that $\!T\partial^-\,X_{\delta
q}^-=X_{\delta q}^-\,\partial^-$.
Equations~\eqref{eq:discrete-LDA} may be written
\[
  \tilde v\in (\partial^-|\kD_d)^{-1}(q),\qquad
  \w{\sigma\-6_d}(\tilde v)\,X_{\delta q}^-(\tilde v)=-\w{\sigma\-6_d}(v)\,
    \delta q,\qquad
  \delta q\in(\kE_d)_v.
\]
The linearization of these in $\tilde v$ is the derivative with
respect to $\tilde v$, in direction $\delta\tilde
v\in\w{\!T\-4_}{\tilde v}\bigl((\partial^-|\kD_d)^{-1}(q)\bigr)$, of
the left side of the second equation. Such $\delta\tilde v$ are
obtained one-to-one as $\delta\tilde q^+$ from $\delta\tilde q\in
\!T\partial^+(\w{\vrt^-_{\tilde v}\-3}\kV\cap\w{\!T\-4_}{\tilde
v}\kD_d)$. So, from Equations~\eqref{eq:def-dpm-dmp}, the required
condition is that the bilinear form
\[[eq:LDA-prime-d-regularity]
  (\delta q,\delta\tilde q)\mapsto\!d^\pm\w{\sigma\-6_d}(\tilde v)(\delta q,
    \delta\tilde q),\qquad
  \delta q\in(\kE_d)_v,\;
    \delta\tilde q\in\!T\partial^+(\w{\vrt^-_{\tilde v}\-3}\kV
    \cap\w{\!T\-4_}{\tilde v}\kD_d)
\]
is nonsingular.

In continuous Lagrangian mechanics, the term `regular' refers to
linear conditions that provide proper equations of motion
(hyperregular is the global condition that the Legendre transform is a
diffeomorphism), and there is a single notion of regular, which is
equivalent to nondegeneracy of the Lagrange two-form. But linear
conditions are of the infinitesimal, and they do not migrate well to
the discrete context, which is finite. The discrete tangent bundle
does not intrinsically support linear operations. So it is not that
surprising to find a variety of notions of regularity in the discrete
context, and we collect some of these here. There are more
possibilities than the below: for example, more can be generated by
replacing $\!d^\pm$ with $\!d^\mp$.

\begin{definition}\label{def:regular-DCLS}
  Let $(\kV,\w{\sigma\-6_d},\kD_d,\kE_d)$ be a DCLS.
  \begin{enumerate}
    \item
      $\w{\sigma\-6_d}$ is \defemph{regular$^{-}$} if, for all
      $(v,\tilde v)\in\ddot\kQ$ such that $v,\tilde v\in\kD_d$, the
      conditions (1)~$\!d^\pm\w{\sigma\-6_d}(\tilde v)(\delta
      q,\delta\tilde q)=0$ for all $\delta\tilde q\in(\kE_d)_{\tilde
        v}$, and (2)~$\delta q\in\!T\partial^-(\w{\vrt^+_{\tilde v}\-3}\kV\cap\w{\!T\-4_}{\tilde v}\kD_d)$, imply $\delta
      q=0$;
    \item
      $\w{\sigma\-6_d}$ is \defemph{regular$^{+}$} if, for all
      $(v,\tilde v)\in\ddot\kQ$ such that $v,\tilde v\in\kD_d$, the
      conditions (1)~$\!d^\pm\w{\sigma\-6_d}(\tilde v)(\delta
      q,\delta\tilde q)=0$ for all $\delta q\in(\kE_d)_v$, and
      (2)~$\delta\tilde q\in\!T\partial^+(\w{\vrt^-_{\tilde v}\-3}\kV\cap\w{\!T\-4_}{\tilde v}\kD_d)$, imply $\delta
      \tilde q=0$;
    \item
      $\w{\sigma\-6_d}$ is \defemph{regular} if it is regular$^{-}$ and
      regular$^{+}$;
    \item
      $\w{\sigma\-6_d}$ is \defemph{$(V,\tilde V)$-regular}, $V,\tilde
      V\subseteq\kD_d$ open, if it is regular, and, for all $v\in V$
      there is a unique $\tilde v\in\tilde V$ such that $(v,\tilde v)$
      satisfies the discrete Lagrange--d'Alembert
      Equations~\eqref{eq:discrete-LDA}.
   \end{enumerate}
\end{definition}
\noindent
The fiber dimension of $\!T\partial^+(\w{\vrt^-\-3\kV}\cap\!T\kD_d)$ is
$\dim\kD_d-\dim\kQ$ and so, if $\w{\sigma\-6_d}$ is regular$^+$, then
necessarily
\[[eq:DCLS-dimension]
  \dim\kD_d-\dim\kQ=\fdim\kE_d,
\]
and, as in the continuous context, we denote the common value
by~$r$. Thus regularity implies dimensional equality of the
constraints and constraint forces, just as in the continuous context.

A \defemph{discrete Lagrangian vector field} is a map
$Y_{\w{\sigma\-6_d}}\-2\colon U\to\ddot\kQ$, where $U\subseteq\kV$ is
open, such that $\partial_{\kV\times\kV}^-\,Y_{\w{\sigma\-6_d}}(v)=v$
and
\[
  \mbox{$\displaystyle\delta\w{\sigma\-6_d}\,Y_{\w{\sigma\-6_d}}(v)
    \,\delta q=0$
    \quad for all\quad
    $\displaystyle\delta q\in(\kE_d)_v$.}
\]
By Theorem~\ref{thm:reduce-to-N=2}, $v_d$ is a discrete evolution if
\[
  \bigl(v_d(k),v_d(k+1)\bigr)=Y_{\w{\sigma\-6_d}}\bigl(v_d(k)\bigr),
\]
which says that the discrete derivative of the sequence $v_d$ at $k$
is the discrete Lagrange vector field at $v_d(k)$. Discrete
evolutions can be obtained from a discrete Lagrangian vector field by
iterations of maps $F$ defined by
$Y_{\w{\sigma\-6_d}}(v)=\bigl(v,F(v)\bigr)$.

%
\section{Structures of discrete Lagrangian systems}
\label{sec:DCLS-structures}
%
%
\noindent Beginning with~\cite{MarsdenJE-PatrickGW-ShkollerS-1998-1},
and continuing with~\cite{PatrickGW-2007-1}, there is an effective
procedure for the recognition of structure for variational theories,
specifically symplecticity, momentum preservation, and the equations
of motion. In summary, this procedure uses the decomposition of the
action into boundary and nonboundary parts, such as
Equations~\eqref{eq:dS-continuous} and~\eqref{eq:dS-discrete}. This
decomposition is pulled back by the inclusion~$\iota$ which maps
solutions into the domain of the action functional. In the context of
a Lagrangian $L$ and action $\w{S\-5_t}$, the procedure consists of
the following steps:

\begin{description}
  \item[{\it Momentum structure:}]
    write $\!i_\xi\bigl(\iota^*\!d\w{S\-5_t}\bigr)=0$;
  \item[{\it Symplectic structure:}]
    write $\!d\bigl(\iota^*(\!d\w{S\-5_t})\bigr)=0$;
  \item[{\it Symplectic equations structure:}]
    note $\displaystyle L=\frac d{dt}\biggr|_{t=0}\iota^*\w{S\-5_t}$,
    write $\displaystyle\!dL=\frac
    d{dt}\biggr|_{t=0}\iota^*\!d\w{S\-5_t}$ and use
    $\displaystyle\!L_X\alpha=\!d\!i_X\alpha+\!i_X\!d\alpha$.
\end{description}
In this section we apply this procedure to extract the discrete
structure preservation properties of a DCLS. \emph{From the remaining
of Section~\ref{sec:DCLS-structures} through
Section~\ref{sec:LdA-skew-critical-problem}, let
$(\kV,\partial^-,\partial^+,\w{\sigma\-6_d},\kD_d,\kE_d)$ be a given
DCLS with be an evolution map $F\colon U_F\to V_F$.}

%
\subsection{Decomposition}
\label{sec:DCLS-decomposition}
%
%
\noindent The evolution map~$F$ defines an insertion $\iota_F$ of $U_F$
into solutions of the DCLS by
$\iota_F(v)\=\bigl(v,F(v)\bigr)$. Pulling back $\Sigma_d$ by $\iota_F$
gives, from Equation~\eqref{eq:dS-discrete},
\[
  \iota_F^*\Sigma_d(v)\,\delta v=
    \delta\w{\sigma\-6_d}\bigl(v,F(v)\bigr)\,\!T\partial^+(\delta v)
    +\theta_{\w{\sigma\-6_d}}^+\bigl(F(v)\bigr)\,\!TF(\delta v)
    -\theta_{\w{\sigma\-6_d}}^-(v)\,\delta v
\]
i.e.
\[[eq:DCLS-Sigma-to-solutions]
  \iota_F^*\Sigma_d=F^*\theta_{\w{\sigma\-6_d}}^+-\theta_{\w{\sigma\-6_d}}^-+
    \w{\alpha\-2_F}
\]
where $\w{\alpha\-2_F}$ is the one form defined by
\[[eq:def-alphaF]
 \w{\alpha\-2_F}(v)\,\delta v\=\delta\w{\sigma\-6_d}\bigl(v,F(v)\bigr)\,
   \!T\partial^+(\delta v).
\]
An important fact is that
\[[eq:alphaF-annihilates]
  \mbox{$\displaystyle\w{\alpha\-2_F}(v)\,\delta v=0$
    \;for all\; $\displaystyle\delta v\in\w{\!T\-4_}v\kD_d$
    \; such that\;
    $\displaystyle\!T\partial^+(\delta v)\in(\kE_d)_v$,}
\]
because $\bigl(v,F(v)\bigr)$ is a solution pair, and because of
Equations~\eqref{eq:discrete-LDA} and~\eqref{eq:def-alphaF}.

%
\subsection{Momentum}
\label{sec:DCLS-momentum}
%
%
\noindent We begin with an definition of a \defemph{symmetric
DCLS}. Let $\kG$ be a Lie group with Lie algebra $\fg$.

\begin{definition}\label{def:symmetric-DCLS}
  An \defemph{action of a group $\kG$ on the DCLS
  $(\kV,\partial^-,\partial^+,\w{\sigma\-6_d},\kD_d,\kE_d)$}, where
  $\partial^-\-2\colon\kV\to\kQ$ and $\partial^+\-2\colon\kV\to\kQ$, means
  actions of $\kG$ on $\kQ$ and $\kV$ such that
  \begin{enumerate}
    \item $\partial^-$ and $\partial^+$ are intertwining; and
    \item $\w{\sigma\-6_d}$, $\kD_d$, $\kE_d$ are invariant; and
    \item $\w{\sigma\-6_d}(v)(\xi v)=0$ for all $\xi\in\fg$ and all $v\in\kD_d$.
  \end{enumerate}
\end{definition}
\noindent
In a symmetric DCLS, the derived constructs $\kC_{d,N}$, $\kN_{d,N}$,
$\kW_{d,N}$, $\Sigma_d$ are all invariant under diagonal actions of
$\kG$, and the symmetry group preserves the solutions. \emph{For the
remainder Section~\ref{sec:DCLS-momentum}, we posit the action of a
group as in Definition~\ref{def:symmetric-DCLS}.}

Equivariance of evolutions requires uniqueness and hence has to rely
on regularity. However, infinitesimal equivariance can be recovered
with only an infinitesimal flavor of regularity.

\begin{theorem}\mbox{}

  \begin{enumerate}
    \item
      If $\w{\sigma\-6_d}$ is regular$^+$ then $\!TF(\xi v)=F(\xi v)$ for all
      $v$ in the domain of $F$ and all $\xi\in\fg$.
    \item
      Suppose $\w{\sigma\-6_d}$ is $(U_F,V_F)$-regular$^+$. Then
      $\!TF(gv)=gF(v)$ for all $v\in U_F$ and $g\in\kG$ such that
      $gv\in V_F$.
  \end{enumerate}
\end{theorem}

\begin{proof}
From  discussion following Equation~\eqref{eq:discrete-LDA},
$\w{\sigma\-6_d}$ regular$^+$ implies local existence and uniqueness of the
evolutions via the inverse function theorem. Thus, for small $t$,
$\kG$ invariance of the evolutions implies $F\bigl(\exp(\xi
t)v\bigr)=\exp(\xi t)\tilde v$ and the first statement is obtained by
differentiation at $t=0$. The second statement follows from the
global existence and uniqueness of $(U_F,V_F)$-regularity.\elsqed
\end{proof}

Generally, momentum is defined by insertion of the infinitesimal
generator into the analogue of the Lagrange one form. There are two
momenta since the DCLS context includes two Lagrange one forms
$\theta_{\w{\sigma\-6_d}}^+$ and $\theta_{\w{\sigma\-6_d}}^-$.

\begin{definition}\label{def:DCLS-momentum}
  The \defemph{momentum maps} are the two functions
  $J^+\-3\colon\kV\rightarrow\fg^*$ and $J^-\-3\colon\kV\rightarrow\fg^*$
  defined by
  \[
    J^-_\xi(v)\=\langle J^-(v),\xi\rangle
      \=\theta_{\w{\sigma\-6_d}}^-(v)(\xi v)=-\w{\sigma\-6_d}(v)(\xi v)^-,\qquad
    J^+_\xi(v)\=\langle J^+(v),\xi\rangle\=
      \theta_{\w{\sigma\-6_d}}^+(v)(\xi v).
  \]
\end{definition}

\noindent
Group invariance provides that both the momentum maps intertwine the
action on $\kV$ and the coadjoint action of $\fg^*$ i.e.\ the momenta
are $\CoAd$-equivariant. Furthermore, if $v\in\kD_d$ then
$J^-(v)=J^+(v)$ and the superscript on $J$ may be dropped, because
\[
  0=\langle\w{\sigma\-6_d}(v),\xi v\rangle
    =\langle\w{\sigma\-6_d}(v),(\xi v)^-\rangle+\langle\w{\sigma\-6_d}(v),
      (\xi v)^+\rangle
    =-J^-_\xi(v)+J^+_\xi(v).
\]

The momentum conservation structure for a DCLS is  as follows:
\begin{theorem}\label{thm:DCLS-momentum-equation}
  If $\w{\sigma\-6_d}$ is regular$^+$, $v\in\kD_d$, and $F$ is an evolution
  map, then $J_\xi\bigl(F(v)\bigr)=J_\xi(v)$ for any $\xi\in\fg$ such
  that $\xi\,\partial^+(v)\in(\kE)_v$.
\end{theorem}

\begin{remark}
  The momenta $J_\xi$ are of course not generally conserved in
  continuous nonholonomic
  mechanics. Theorem~\ref{thm:DCLS-momentum-equation} does not imply
  conservation of arbitrary momenta for a DCLS, because if
  $v_d\in\kV[1,N]$ is an evolution then it is not usually possible to
  arrange the condition
  $\xi\,\partial^+\bigl(v_d(k)\bigr)\in(\kE)_{v_d(k)}$ for constant
  $\xi$ independent of $k$. Rather, one will have a map $\kV\ni
  v\rightarrow\xi^\kV(v)$ which satisfies
  $\xi^\kV(v)\,\partial^+(v)\in(\kE)_v$, and then
  \[
    J_{\xi^\kV(F(v))}\bigl(F(v)\bigr)-J_{\xi^\kV(v)}\bigl(v\bigr)
      =J_{\xi^\kV(F(v))}(v)-J_{\xi^\kV(v)}(v)
      =\bigl\langle J(v),\xi^\kV\bigl(F(v)\bigr)
      -\xi^\kV\bigl(v\bigr)\bigr\rangle,
  \]
  which is called the discrete nonholonomic momentum
  equation~\cite{CortesJ-MartinezS-2001-1}.
\end{remark}

\begin{proof}[Proof of Theorem~\ref{thm:DCLS-momentum-equation}]
Insertion of the infinitesimal generator $\xi_{\kD_d}$ into
Equation~\eqref{eq:DCLS-Sigma-to-solutions} gives, because $\w{\sigma\-6_d}$
annihilates infinitesimal generators,
\[
  0=\bigl\langle\theta^+_{\w{\sigma\-6_d}}\bigl(F(v)\bigr),\xi\,F(v)\bigr\rangle
    -\bigl\langle\theta^-_{\w{\sigma\-6_d}}(v),\xi v\bigr\rangle
    =J^+_\xi\bigl(F(v)\bigr)-J^-_\xi\bigl(v\bigr).
\]
There is no contribution from $\w{\alpha\-2_F}$ because of the
Equation~\eqref{eq:alphaF-annihilates}.\elsqed
\end{proof}

%
\subsection{Symplectic}
\label{sec:DCLS-symplectic}
%
%
\noindent In the continuous nonholonomic systems \LdA\ with constraint
$\kD$ and variations $\kE$, the distribution
\[
  \kK_{\kD,\kE}\=\set{\delta v\in\w{\!T\-4_}v\kD}{\!T\w{\tau\-2_\kQ}
    (\delta v)\in\kE_v}
\]
is an important object because it supports the associated nonholonomic
semi-symplectic
structure~\cite{BatesL-SniatyckiJ-1993-1,
                CushmannR-KemppainenD-SniatyckiJ-BatesL-1995-1,
                PatrickGW-2007-1,
                SniatyckiJ-2002-1}.
For a DCLS we
explore three possible discrete analogues of $\kK_{\kD,\kE}$:
\[[eq:DCLS-K-distributions]
  &\kK_{\kD_d,\kE_d}^-\=\set{\delta v\in\w{\!T\-4_}v\kD_d}
    {\!T\partial^-(\delta v)\in(\kE_d)_{\partial^-(v)}},
  \\
  &\kK_{\kD_d,\kE_d}^0\=\set{\delta v\in\w{\!T\-4_}v\kV}
    {\!T\partial^-(\delta v)\in(\kE_d)_{\partial^-(v)},
    \!T\partial^+(\delta v)\in(\kE_d)_{\partial^+(v)}},
  \\
  &\kK_{\kD_d,\kE_d}^+\=\set{\delta v\in\w{\!T\-4_}v\kD_d}
    {\!T\partial^+(\delta v)\in(\kE_d)_{\partial^+(v)}}.
\]
Under a the dimension condition~\eqref{eq:DCLS-dimension}, which is
implied by regularity, all three of these discrete analogues have the
same fiber dimension as the continuous $\kK_{\kD,\kE}$:

\begin{lemma}\label{lem:fdim-KDds}
  If $r\=\dim\kD_d-\dim\kQ=\fdim\kE_d$ then
  \[
    \fdim\kK_{\kD_d,\kE_d}^0=\fdim\kK_{\kD_d,\kE_d}^+=\fdim\kK_{\kD_d,\kE_d}^-=2r.
  \]
\end{lemma}

\begin{proof}
By definition, $\partial^+|\kD_d$ is a submersion, so
\[
  \dim\kQ+r-\fdim\kK_{\kD_d,\kE_d}^+=\dim\kQ-\fdim\kE,
\]
so $\fdim\kK_{\kD_d,\kE_d}^+=2r$ follows and
$\fdim\kK_{\kD_d,\kE_d}^-=2r$ is similar. The fibers of
$(\!T\partial^-)^{-1}\kE_d$ and $(\!T\partial^-)^{-1}\kE_d$ are
transversal subspaces of the fibers of $\!T\kV$, since
\[
  \w{\vrt^+\-3\kV}=\ker\!T\partial^+\subseteq(\!T\partial^+ )^{-1}\kE_d,\qquad
  \w{\vrt^-\-3\kV}=\ker\!T\partial^-\subseteq(\!T\partial^- )^{-1}\kE_d,
\]
and $\vrt^+V\oplus\w{\vrt^-\-3\kV}=\!T\kV$. The codimension of the intersection
of transversal subspaces is the sum of the codimensions, so
\[
  \cofdim\kK_{\kD_d,\kE_d}^0=\cofdim(\!T\partial^+ )^{-1}\kE_d
    +\cofdim(\!T\partial^- )^{-1}\kE_d=2(\dim\kQ-r)
\]
and the result follows because
\begin{equation*}
  \cofdim\kK_{\kD_d,\kE_d}^0=\dim\kV-\fdim\kK_{\kD_d,\kE_d}^0
    =2\dim\kQ-\fdim\kK_{\kD_d,\kE_d}^0.
  \tag*{\elsqed}
\end{equation*}
\end{proof}

The distribution $\kK_{\kD_d,\kE_d}^+$ is the kernel of the canonical
vector bundle mapping $\nu\colon
\!T\kV\rightarrow\!T\kV/\kK_{\kD_d,\kE_d}^+$ and
admits~\cite{PatrickGW-2007-1} a \defemph{curvature two form}
$\Delta_{\kK_{\kD_d,\kE_d}^+}$ on $\!T\kV$ such that, for all vector
fields $X,Y\in{\kK_{\kD_d,\kE_d}^+}$,
\[
  \Delta_{\kK_{\kD_d,\kE_d}^+}(X,Y)\=-\nu[X,Y].
\]
Clearly $\kK_{\kD_d,\kE_d}^+$ is involutive if and only if
$\Delta_{\kK_{\kD_d,\kE_d}^+}=0$.

\begin{theorem}\label{thm:DCLS-symplectic}
  Let $\kK_d^0$ be a subbundle of $\kK_{\kD_d,\kE_d}^+$ over
  $\kD_d^0\subseteq\kD_d$. Suppose that
  \begin{enumerate}
    \item\label{enum:DCLS-symplectic-1}$\kK_d^0$ is $\!TF$ invariant;
    \item\label{enum:DCLS-symplectic-2} $\Delta_{\kK_{\kD_d,\kE_d}^+}\-9=0$
      on $\kK_d^0$;
    \item\label{enum:DCLS-symplectic-3} $\!d\w{\sigma\-6_d}=0$ on $\kK_d^0$.
  \end{enumerate}
  Then $F$ preserves
  $\omega_{\w{\sigma\-6_d}}=\omega^+_{\w{\sigma\-6_d}}=\omega^-_{\w{\sigma\-6_d}}$ on
  $\kK_d^0$.
\end{theorem}

\begin{proof}
If $\delta v,\delta w\in(\kK_d^0)_v$ then, remembering
the inclusion $\iota_F(v)\=\bigl(v,F(v)\bigr)$,
\[
  \!d(\iota_F^*\Sigma_d)(\delta v,\delta w)
    =\iota_F^*\,\!d\Sigma_d(\delta v,\delta w)
    =\!d\w{\sigma\-6_d}(v)(\delta v,\delta w)
      +\!d\w{\sigma\-6_d}\bigl(F(v)\bigr)\bigl(\!TF(\delta v),
      \!TF(\delta w)\bigr)=0,
\]
because of Items~\ref{enum:DCLS-symplectic-1}
and~\ref{enum:DCLS-symplectic-3}. In the same way,
$\w{\sigma\-6_d}=\theta_{\w{\sigma\-6_d}}^+-\theta_{\w{\sigma\-6_d}}^-$
and $\!d\w{\sigma\-6_d}=0$ on $\kK_d^0$ imply
$\omega^+_{\w{\sigma\-6_d}}=-\!d\theta_{\w{\sigma\-6_d}}^+
=\!d\theta_{\w{\sigma\-6_d}}^-
=\omega^-_{\w{\sigma\-6_d}}$ on $\kK_d^0$. Thus, from the exterior
derivative of Equation~\eqref{eq:DCLS-Sigma-to-solutions},
\[
  F^*\omega_{\w{\sigma\-6_d}}^+=\!d\omega_{\w{\sigma\-6_d}}^-+\!d\w{\alpha\-2_F},
\]
on $\kK_d^0$, so it is sufficient to show $\!d\w{\alpha\-2_F}=0$ on $\kK_d^0$.
Extending $\delta v$ and $\delta w$ to vector fields $V\in\kK_d^0$ and
$W\in\kK_d^0$,
\[
  \!d\w{\alpha\-2_F}(\delta v,\delta w)=V\bigl(\w{\alpha\-2_F}(W)\bigr)(v)
    -W\bigl(\w{\alpha\-2_F}(V)\bigr)(v)-\w{\alpha\-2_F}([V,W])(v),
\]
and the result follows from Equation~\eqref{eq:alphaF-annihilates}
because Item~\ref{enum:DCLS-symplectic-2} implies
$[V,W](v)\in\kK_{\kD_d,\kE_d}^+$.\elsqed
\end{proof}

%
\section{Discrete linear and holonomic constraints}
\label{sec:discrete-holonomic}
%
%
\noindent The continuous constrained Lagrangian systems \LdA\ commonly
have linear constraints, where the constraint is a distribution on
$\kQ$, and the same distribution also provides the variations i.e. the
special case where $\kD$ is a distribution and
$\kE\=(\w{\tau\-2_\kQ}|\kD)^*\kD$ is the usual one. In this section we
construct discrete analogues of this special case.

Recall~\cite{SussmannHJ-1973-1} that if $m,\tilde m\in\kM$, and $\kF$
is a distribution on $\kM$, then $\tilde m$ is
\defemph{$\kF$-reachable} from $m$ if there is a piecewise smooth
curve $c\colon[a,b]\to\kQ$ such that $c^\prime(t)\in\kF$ and $c(a)=m$,
$c(b)=\tilde m$. $\kF$-reachability is an equivalence relation on
$\kM$ and the equivalence classes are called the \defemph{orbits}
of~$\kF$.

\begin{definition}\label{def:classical-DCLS}
  The DCLS $(\kV,\partial^+,\partial^-,\w{\sigma\-6_d},\kD_d,\kE_d)$
  \defemph{has linear constraints} if there is a distribution
  $\w{\kD_{\-6d}^\kQ}$ on $\kQ$ such that
  \begin{enumerate}
    \item
      $(\kE_d)_v=(\w{\kD_{\-6d}^\kQ})_{\partial^+(v)}$ for all $v\in\kD_d$; and
    \item\label{enum:def:classical-DCLS-2}
      $\partial^-(v)$ and $\partial^+(v)$ are $\w{\kD_{\-6d}^\kQ}$-reachable
      for all  $v\in\kD_d$; and
    \item
      $\dim\kD_d=\dim\w{\kD_{\-6d}^\kQ}$.
  \end{enumerate}
  $(\kV,\partial^+,\partial^-,\w{\sigma\-6_d},\kD_d,\kE_d)$
  is \defemph{holonomic} if $\w{\kD_{\-6d}^\kQ}$ is involutive and
  $\!d\w{\sigma\-6_d}=0$.
\end{definition}
\noindent To compare this definition with the continuous context, the
first condition corresponds to using the distribution $\kD^\kQ$ of
$\kQ$ for the variations~$\kE$. The second condition is fulfilled, for
example, in the case where the tangent bundle $\kV$ arises from a
discretization which curve segments are integral curves of
$\w{\kD_{\-6d}^\kQ}$. Such curves can be regarded as discrete
analogues of the elements of $\w{\kD_{\-6d}^\kQ}$, so the second
condition corresponds to equating $\w{\kD_{\-6d}^\kQ}$ with the
velocity constraint. Thus a DCLS with linear constraints is a discrete
analogue of a (continuous) \LdA\ with linear constraints, in that a
single distribution on configuration space generates both the velocity
constraint and the variations.

\begin{remark}\label{rem:linear-DCLS-dimension-remark}
  In the context of Definition~\ref{def:classical-DCLS} we will set
  $r\=\fdim\kE_d=\fdim\w{\kD_{\-6d}^\kQ}$. In
  particular,~$\dim\kD_d=\dim\kQ+r$.
\end{remark}

\emph{For the remainder of Section~\ref{sec:discrete-holonomic}, let
the DCLS $(\kV,\partial^+,\partial^-,L_d,\w{\sigma\-6_d},\kE_d)$
have linear constraint distribution~$\w{\kD_{\-6d}^\kQ}$.} If
$\w{\kD_{\-6d}^\kQ}$ is integrable, then the condition that
$\partial^-(v)$ and $\partial^+(v)$ are $\kD_d^\kQ$-reachable is
strong, because it confines these to be in the same $r$-dimensional
leaf of $\w{\kD_{\-6d}^\kQ}$. Lemma~\ref{lem:involutive-DQ} is a first
step in this line of reasoning.

\begin{lemma}\label{lem:involutive-DQ}
  Suppose $\w{\kD_{\-6d}^\kQ}$ is involutive and let $q\in\kQ$. Then
  $\partial^+$ and $\partial^-$ are local diffeomorphisms from
  (respectively) $\w{\kV^-_{\-5q}}\cap\kD_d$ and
  $\w{\kV^+_{\-5q}}\cap\kD_d$ to the leaf of $\w{\kD_{\-6d}^\kQ}$
  through $q$.
\end{lemma}

\begin{proof}
Let $\kL_q$ be the leaf of $\w{\kD_{\-6d}^\kQ}$ containing
$q\in\kQ$. By Item~\ref{enum:def:classical-DCLS-2} of
Definition~\ref{def:classical-DCLS}, $\partial^+$ immerses the
$r$-dimensional submanifold $\w{\kV^-_{\-5q}}\cap\kD_d$ into $\kL_q$,
which also has dimension $r$. Thus $\partial^+$ is a local
diffeomorphism from $\w{\kV^-_{\-5q}}\cap\kD_d$ into
$\kL_q$. Similarly, $\partial^-$ a local diffeomorphism from
$\w{\kV^+_{\-5q}}\cap\kD_d$ into $\kL_q$.\elsqed
\end{proof}

\begin{lemma}\label{lem:regular-discrete-linear-constraints}
  If $\w{\kD_{\-6d}^\kQ}$ is involutive then
  \begin{enumerate}
    \item
      $\w{\sigma\-6_d}$ is regular$^-$ if and only if for all
      $v\in\kD_d$, the
      conditions~(1)~$\displaystyle\!d^\pm\w{\sigma\-6_d}(v)(\delta
      q,\delta \tilde q)=0$ for all $\displaystyle\delta \tilde
      q\in(\w{\kD_{\-6d}^\kQ})_{\partial^+(v)}$,
      and~(2)~$\displaystyle\delta
      q\in(\w{\kD_{\-6d}^\kQ})_{\partial^-(v)}$, imply $\delta q=0$.
    \item
      $\w{\sigma\-6_d}$ is regular$^+$ if and only if for all
      $v\in\kD_d$, the
      conditions~(1)~$\displaystyle\!d^\pm\w{\sigma\-6_d}(v)(\delta
      q,\delta\tilde q)=0$ for all $\displaystyle\delta
      q\in(\w{\kD_{\-6d}^\kQ})_{\partial^-(v)}$,
      and~(2)~$\displaystyle\delta\tilde
      q\in(\w{\kD_{\-6d}^\kQ})_{\partial^+(v)}$, imply $\delta\tilde
      q=0$.
  \end{enumerate}
\end{lemma}

\begin{proof}
If $v,\tilde v\in\kD_d$ and $\partial^+(v)=\partial^-(\tilde v)$, then
$(\kE_d)_v=(\w{\kD_{\-6d}^\kQ})_{\partial^-(\tilde v)}$.
Lemma~\ref{lem:regular-discrete-linear-constraints} gives
$\!T\partial^+(\w{\vrt^-_{\tilde v}\-3}\kV\cap\w{\!T\-4_}{\tilde
  v}\kD_d) =(\w{\kD_{\-6d}^\kQ})_{\partial^+(\tilde v)}$, and the
result for regular$^+$ is obtained by transcribing
Definition~\ref{def:regular-DCLS}; regular$^-$ is similar.\elsqed
\end{proof}

A skew critical problem in the meaning of
Remark~\ref{rem:skew-ordinary-definition}
is \defemph{ordinary} \defemph{[resp.\ variational]} if the
tangent bundle of the constraint is the equal to [resp.\ contains] the
distribution used to differentiate the objective. Ordinary critical
problems correspond to the standard constrained optimization problem
that seeks critical points of an objective subject to a
constraint. Continuous systems with linear constraints are are
variational exactly if the constraint distribution $\kD$ is integrable
i.e.\ exactly if the system is holonomic in the usual meaning of the
term~\cite{PatrickGW-2007-1}. In the discrete context there is
Theorem~\ref{thm:LdA-holonomic-equivalents} below.

\begin{theorem}\label{thm:LdA-holonomic-equivalents}
  The following are equivalent:
  \begin{enumerate}
    \item
      $\w{\kD_{\-6d}^\kQ}$ is involutive;
    \item
      $\kW_{d,N}$ is involutive;
    \item\label{enum:LdA-holonomic-equivalents-3}
      $\kW_{d,N}=\set{\delta v_d\in\w{\!T\-2\kN}_{d,N}}{\delta q^-(1)=0,\;
        \delta q^+(N)=0}$.
    \end{enumerate}
\end{theorem}

\begin{proof} Assume $N=2$; the proof for arbitrary $N$ is similar.

\smallskip\noindent(1)$\Rightarrow$(3).
Let $(v,\tilde v)\in\kN_d$, and define
\[
  q^-\=\partial^-(v),\qquad
  q\=\partial^+(v)=\partial^-(\tilde v),\qquad
  \tilde q^+\=\partial^+(\tilde v),
\]
which are all in the same leaf of $\w{\kD_{\-6d}^\kQ}$. Temporarily define
\[
  (\dot\kN_d)_{(v,\tilde v)}\=\set{(\delta v,\delta\tilde v)\in
    \w{\!T\-4_}{(v,\tilde v)}(\kV\times\kV)}{
    \!T\partial^-(\delta v)=0,\,\!T\partial^+(\delta\tilde v)=0,\,
    \!T\partial^+(\delta v)=\!T\partial^-(\delta\tilde v),\,
    \delta v\in\!T\kD_d,\,\delta\tilde v\in\!T\kD_d}.
\]
corresponding to the set on the right side of the equality in
Item~\ref{enum:LdA-holonomic-equivalents-3}. Recall that
\[
  (\kW_d)_{(v,\tilde v)}\=\set{(\delta v,\delta\tilde v)\in
    \w{\!T\-4_}{(v,\tilde v)}\kC_d}{\!T\partial^-(\delta v)=0,\,
    \!T\partial^+(\delta\tilde v)=0,\,
    \!T\partial^+(\delta v)=\!T\partial^-(\delta\tilde v)\in\w{\kD_{\-6d}^\kQ}}.
\]
It is required to show that $\kW_{(v,\tilde v)}=\dot\kN_{(v,\tilde
  v)}$ i.e.\ the condition $\delta v,\delta\tilde v\in\!T\kD_d$ in the
definition of $(\kW_d)_{(v,\tilde v)}$ amounts to the same thing as
the condition $\!T\partial^+(\delta v)=\!T\partial^-(\delta\tilde
v)\in\w{\kD_{\-6d}^\kQ}$ in the definition of $\dot\kN_{(v,\tilde v)}$.  If
$(\delta v,\delta\tilde v)\in\dot\kN_{(v,\tilde v)}$ then $\delta v$
is tangent to $\w{\kV^-_{\-5q}}\cap\kD_d$ and $\w{\!T\-4_}v\partial^+$
maps the tangent space of this at $v$ (isomorphically) to
$(\w{\kD_{\-6d}^\kQ})_q$. Thus $\!T\partial^+(\delta
v)\in\w{\kD_{\-6d}^\kQ}$. On the other hand, if $(\delta
v,\delta\tilde v)\in(\kW_d)_{(v,\tilde v)}$, find the unique $\delta
v^\prime\in\w{\vrt^-_v\-3}\kV\cap\kD_d$ and $\delta\tilde
v^\prime\in\w{\vrt^+_v\-3}\kV\cap\kD_d$ such that
$\!T\partial^+(\delta v^\prime)=\!T\partial^-(\delta\tilde
v^\prime)$. This implies $\delta v=\delta v^\prime$ since
$\w{\!T\-4_}v\partial^+$ is a local diffeomorphism from
$\w{\vrt^-_v\-3}\kV$ to $\w{\!T\-4_}qQ$. Thus $\delta v\in
\w{\!T\-4_}v\kD_d$ since $\delta v^\prime$ is. Similarly,
$\delta\tilde v\in\w{\!T\-4_}{\tilde v}\kD_d$, so $(\delta
v,\delta\tilde v)\in\dot\kN_{(v,\tilde v)}$.

\smallskip\noindent(2)$\Rightarrow$(1). Let $X$ and $Y$ be vector
fields with values in $\w{\kD_{\-6d}^\kQ}$ and let $q\in\kQ$. Arrange
$v,\tilde v\in\kN_d$ so that $\partial^+(v)=\partial^-(\tilde v)=q$.
Define $\pi_d\colon\kN_d\rightarrow\kQ$ by $\pi_d(v,\tilde
v)\=\partial^+(v) =\partial^-(\tilde v)$.  $X$ and $Y$ have have
unique lifts $\tilde X$ and $\tilde Y$ which are vector fields on
$\kN_d$ with values in $\kW_d$ such that $\!T\pi_d\,\tilde
X=X\circ\pi_d$ and $\!T\pi_d\,\tilde Y=Y\circ\pi_d$. Then
$[X,Y]\circ\pi_d=\!T\pi_d\,[\tilde X,\tilde Y]$ and by assumption
$[\tilde X,\tilde Y]$ has values in $\kW_d$, so
\[\relax
  [X,Y](q)=\!T\pi_d[\tilde X,\tilde Y](v,\tilde v)\in\w{\kD_{\-6d}^\kQ},
\]
as required.

\smallskip\noindent(3)$\Rightarrow$(2). By assumption, $\kW_d$ is
integrable because it is equal to the kernel of the derivative of the
map from $\kN_d$ to $\kQ\times\kQ$ defined by
$v_d\mapsto\bigl(\partial^-(v),\partial^+(\tilde v)\bigr)$, and any
such kernel is involutive.\elsqed
\end{proof}

There are the following specializations of the
distributions~\eqref{eq:DCLS-K-distributions} to the context with
linear constraints:
\[
  &\kK_{\kD_d}^-\=\!T\kD_d\cap(\!T\partial^-)^{-1}\w{\kD_{\-6d}^\kQ},
  \\
  &\kK_{\kD_d}^0\=\bigl((\!T\partial^-)^{-1}\w{\kD_{\-6d}^\kQ}\cap
    (\!T\partial^+)^{-1}\w{\kD_{\-6d}^\kQ}\bigr)\bigr|_{\kD_d},
  \\
  &\kK_{\kD_d}^+\=\!T\kD_d\cap(\!T\partial^+)^{-1}\w{\kD_{\-6d}^\kQ}.
\]
The fiber dimensions of these are all $2r$ because the dimension
condition of Lemma~\ref{lem:fdim-KDds} is satisfied, as explained in
Remark~\ref{rem:linear-DCLS-dimension-remark}.
Lemma~\ref{lem:kDd-all-collapse} below shows simplifications if
$\w{\kD_{\-6d}^\kQ}$ is involutive, and it provides a discrete
analogue of the continuous fact that $\kK_D$ is integrable if and only
if $\kD$ is integrable.

\begin{lemma}\label{lem:kDd-all-collapse}\mbox{}
  \begin{enumerate}
  \item
    The following are equivalent:
      (1A)~$\w{\kD_{\-6d}^\kQ}$ is involutive,
      (1B)~$\kK_{\kD_d}^-$ is involutive,
      (1C)~$\kK_{\kD_d}^0$ is involutive,
      (1D)~$\kK_{\kD_d}^+$ is involutive.
  \item
    The following are equivalent:
      (2A)~$\kK_{\kD_d}^0\subseteq\!T\kD_d$,
      (2B)~$\kK_{\kD_d}^0=\kK_{\kD_d}^-$,
      (2C)~$\kK_{\kD_d}^0=\kK_{\kD_d}^+$.
  \end{enumerate}
  Moreover, the statements in (1) imply the statements in (2).
\end{lemma}

\begin{proof}\mbox{}

\smallskip\noindent
(1A)$\Rightarrow$(1B), (1A)$\Rightarrow$(1C), (1A)$\Rightarrow$(1D): pull
backs and intersections of involutive distributions are involutive.

\smallskip\noindent
(1A)$\Leftarrow$(1B), (1A)$\Leftarrow$(1C), (1A)$\Leftarrow$(1D): If
$\kK_{\kD_d}^0$ is involutive and $X,Y$ are vector fields on $\kQ$
with $X,Y\in\w{\kD_{\-6d}^\kQ}$, then $X^+,Y^+\in\kK_{\kD_d}^0$, hence
$[X^+,Y^+]\circ\partial^+=\!T\partial^+[X^+,Y^+]\in\w{\kD_{\-6d}^\kQ}$. Thus
$\w{\kD_{\-6d}^\kQ}$ is involutive. (1A)$\Leftarrow$(1B) and
(1A)$\Leftarrow$(1D) are similar after using
Lemma~\ref{lem:involutive-DQ} to lift $X$ and $Y$.

\smallskip\noindent
(2A)$\Leftrightarrow$(2B) and (2A)$\Leftrightarrow$(2C):
If $\kK_{\kD_d}^0\subseteq\!T\kD_d$ then $\kK_{\kD_d}^0\subseteq
\!T\kD_d\cap(\!T\partial^-)^{-1}\w{\kD_{\-6d}^\kQ}=\kK_{\kD_d}^-$ and hence
$\kK_{\kD_d}^0=\kK_{\kD_d}^-$ by Lemma~\ref{lem:fdim-KDds}. If
$\kK_{\kD_d}^0=\kK_{\kD_d}^-$ then
$\kK_{\kD_d}^0\subseteq\kK_{\kD_d}^-\subseteq\!T\kD_d$. Similarly
(2A)$\Leftrightarrow$(2C).

\smallskip\noindent
(1A)$\Rightarrow$(2A): Suppose $\kD^Q_d$ is involutive and $\delta
v\in\kK_{\kD_d}^0$. Then $\!T\partial^+(\delta v^+)=\!T\partial^+(\delta
v)\in\w{\kD_{\-6d}^\kQ}$ and $\delta v^+\in\w{\vrt^-\-3\kV}$. By
Lemma~\ref{lem:involutive-DQ}, there is a $\delta v^\prime\in
\!T(\w{\kV^-_{\-5q}}\cap\kD_d)$ such that $\!T\partial^+(\delta
v^\prime)=\!T\partial^+(\delta v)$. Also $\!T\partial^-(\delta
v)=\!T\partial^-(\delta v^\prime)=0$, so $\delta v=\delta v^\prime\in
\!T\kD_d$.\elsqed
\end{proof}

In a (continuous) nonholonomic system with linear constraints, the
Lagrange two form is nonsingular on the distribution $\kK_\kD$ if and
only if the Lagrangian is regular. For a DCLS, there is the following
similar result in the holonomic case.

\begin{theorem}\label{thm:holonomic-regularity}
  If $(\kV,\partial^+,\partial^-,L_d,\w{\sigma\-6_d},\kE_d)$ is holonomic
  then $\w{\sigma\-6_d}$ is regular$^\pm$ if and only if $\omega_{\w{\sigma\-6_d}}$
  is nondegenerate on $\kK_{\kD_d}^0$.
\end{theorem}

\begin{proof}
Let $v\in\kD_d$ and $\delta v,\delta w\in\kK_{\kD_d}^0$. Then
$\!T\partial^-(\delta v)\in\w{\kD_{\-6d}^\kQ}$, so there is a vector field
$V\in\w{\kD_{\-6d}^\kQ}$ such that $\!T\partial^-(\delta
v)=V\bigl(\partial^-(v)\bigr)$. Similarly choose vector fields
$\tilde V$, $W$, and $\tilde W$ such that
\[
  \!T\partial^+(\delta v)=\tilde V\bigl(\partial^+(v)\bigr),\qquad
  \!T\partial^-(\delta w)=W\bigl(\partial^-(w)\bigr),\qquad
  \!T\partial^+(\delta w)=\tilde W\bigl(\partial^+(w)\bigr).
\]
Then $\delta v=\tilde V^+(v)+V^-(v)$, $\delta w=\tilde W^+(v)+W^-(v)$.
Also,
\[
  &\omega_{\w{\sigma\-6_d}}(\tilde V^++V^-,\tilde W^++W^-)\\
  &\qquad=-(\tilde V^++V^-)\bigl(\theta_{\w{\sigma\-6_d}}^+(\tilde W^++W^-)\bigr)
    +(\tilde W^++W^-)\bigl(\theta_{\w{\sigma\-6_d}}^+(\tilde V^++V^-)\bigr)
    +\theta_{\w{\sigma\-6_d}}^+([\tilde V^++V^-,\tilde W^++W^-])
  \\
  &\qquad=-(\tilde V^++V^-)\bigl(\w{\sigma\-6_d}(\tilde W^++W^-)^+\bigr)
    +(\tilde W^++W^-)\bigl(\w{\sigma\-6_d}(\tilde V^++V^-)^+\bigr)
    +\w{\sigma\-6_d}([\tilde V^++V^-,\tilde W^++W^-]^+)
  \\
  &\qquad=-V^-\bigl(\w{\sigma\-6_d}(\tilde W^+)\bigr)+(W^-)\bigl(\w{\sigma\-6_d}
    (\tilde V^+)\bigr)-\!d\w{\sigma\-6_d}(\tilde V^+,\tilde W^+)
  \\
  &\qquad=\!d^\mp\w{\sigma\-6_d}(W,\tilde V)-\!d^\mp\w{\sigma\-6_d}(V,\tilde W)
\]
so
\[
  \omega_{\w{\sigma\-6_d}}(\delta v,\delta w)
    =\!d^\pm\w{\sigma\-6_d}(v)\bigl(\!T\partial^-(\delta w),\!T\partial^+
    (\delta v)\bigr)
    -\!d^\pm\w{\sigma\-6_d}(v)\bigl(\!T\partial^-(\delta v),\!T\partial^+
    (\delta w)\bigr).
\]
Suppose $\w{\sigma\-6_d}$ is regular$^\pm$ and that
$\omega_{\w{\sigma\-6_d}}(\delta v,\delta w)=0$ for all $\delta
w\in(\kK_d^0)_v$. Choosing $\delta w\in\w{\vrt^-\-3\kV}$ gives
\[
  \!d^\pm\w{\sigma\-6_d}(v)\bigl(\!T\partial^-(\delta v),\delta\tilde q\bigr)=0
\]
for all $\delta\tilde q\in(\w{\kD_{\-6d}^\kQ})_{\partial^+(v)}$. Thus
$\!T\partial^-(\delta v)=0$ as $\w{\sigma\-6_d}$ is regular$^-$. Similarly
$\!T\partial^+(\delta v)=0$ since $\w{\sigma\-6_d}$ is regular$^+$. Thus
$\delta v=0$, proving that $\omega_{\w{\sigma\-6_d}}$ is nondegenerate. The
converse~---~that $\w{\sigma\-6_d}$ is regular$^\pm$ if $\omega_d$ is
nondegenerate on $\kK_{\kD_d}^0$, follows by reversing this augment.\elsqed
\end{proof}

In the holonomic case, symplecticity and preservation of momentum are
expected and they are recovered in the following two corollaries.

\begin{corollary}
  If $F\colon U_F\subseteq\kV\to V_F\subseteq\kV$ is an evolution of
  the holonomic DCLS
  $(\kV,\partial^+,\partial^-,\w{\sigma\-6_d},\kD_d,\kE_d)$, then $F$
  is symplectic on $\kK_{\kD_d}^0$ i.e.\ $\kK_{\kD_d}^0$ is $\!TF$
  invariant and $\omega_d(v)\bigl(\!TF(\delta v),\!TF(\delta
  w)\bigr)=\omega_d(v)(\delta v,\delta w)$ for all $\delta v,\delta
  w\in(\kK_{\kD_d}^0)_v$, $v\in\kD_d$.
\end{corollary}

\begin{proof}
If suffices to verify the hypotheses of
Theorem~\ref{thm:DCLS-symplectic} with $\kK_d^0=\kK_{\kD_d}^0$. If
$v\in\kD_d$ and $\delta v\in(\kK_{\kD_d}^0)_v$ then $\delta v\in
\!T\kD_d$ and $\!T\partial^+(\delta v)\in
(\kE_d)_v=(\w{\kD_{\-6d}^\kQ})_{\partial^+(v)}$. Since $\kD_d$ is $F$
invariant, $\!TF(\delta v)\in\!T\kD_d$. Also
\[
  \!T\partial^-\bigl(\!TF(\delta v)\bigr)
    =\!T(\partial^-\circ F)(\delta v)
    =\!T\partial^+(\delta v)\in\kD_d^Q
\]
so $\!T\partial^-\bigl(\!TF(\delta
v)\bigr)\in\kK_{\kD_d}^-=\kK_{\kD_d}^0$. Furthermore
$\Delta_{\kK_{\kD_d}^+}=0$ because Lemma~\ref{lem:kDd-all-collapse}
implies $\kK_{\kD_d}^+$ is involutive, and $\!d\w{\sigma\-6_d}=0$ by
hypothesis.\elsqed
\end{proof}

\begin{corollary}\label{cor:discrete-holonomic-momentum}
Suppose $F\colon U_F\subseteq\kV\to V_F\subseteq\kV$ is an evolution
of the holonomic DCLS
$\kV\=(\kV,\partial^+,\partial^-,\w{\sigma\-6_d},\kD_d,\kE_d)$, and
suppose that a Lie group $\kG$ acts on $\kV$ and $\kQ$ such that
\begin{enumerate}
  \item\label{enum:discrete-holonomic-momentum-1}
    $\partial^+$ and $\partial^-$ are equivariant; and
  \item\label{enum:discrete-holonomic-momentum-2}
    $\w{\kD_{\-6d}^\kQ}$ is $\kG$ invariant; and
  \item\label{enum:discrete-holonomic-momentum-3}
    $\fg\kQ\subseteq\w{\kD_{\-6d}^\kQ}$.
\end{enumerate}
Then $\kV$ is a symmetric DCLS and $F$ is momentum conserving.
\end{corollary}

\begin{proof}
Invariance of $\kE_d$ follows from
$(\kE_d)_v\=(\w{\kD_{\-6d}^\kQ})_{\partial^+(v)}$ and
Items~\ref{enum:discrete-holonomic-momentum-1}
and~\ref{enum:discrete-holonomic-momentum-2}. From
Item~\ref{enum:discrete-holonomic-momentum-3}, the definition of
$\kK_{\kD_d}^0$, and Lemma~\ref{lem:kDd-all-collapse}, we have
$\fg\kD_d\subseteq\kK_{\kD_d}^0\subseteq\!T\kD_d$, so $\kD_d$ is
invariant and hence $\kV$ is symmetric. By
Theorem~\ref{thm:DCLS-momentum-equation}, conservation of momentum
follows from $\xi\,\partial^+(v)\in(\kE)_v$ for all $v\in\kD_d$ and
all $\xi\in\fg$, but this is implied by
Item~\ref{enum:discrete-holonomic-momentum-3}.\elsqed
\end{proof}

%
\section{Lagrange-d'Alembert as a skew critical problem}
\label{sec:LdA-skew-critical-problem}
%
%
\noindent From~\cite{CuellC-PatrickGW-2007-1} we recover some basic
definitions:

\begin{definition}\label{def:skew-critical-problem}
  Let $M$ and $N$ be manifolds, $\alpha$ be a one-form on $M$, $\kD$
  be a $C^k$ distribution on $M$, and let $g\colon M\to N$ be a
  submersion. We call $(\alpha,\kD,g)$ a \defemph{$C^k$ skew critical
  problem}. A point $m_c\in M$ is a \defemph{skew critical point of
  $(\alpha,\kD,g)$} at $n\in N$ if
  \[
    \left\{\begin{array}{l}
      \mbox{$\displaystyle\alpha(m_c)(v)=0$ for all
      $\displaystyle v\in\kD_{m_c}$,}\\[2pt]\displaystyle g(m_c)=n.
    \end{array}\right.
  \]
  A skew critical problem is called \defemph{variational} if $\kD$ is
  involutive.
\end{definition}

\begin{definition}\label{def:skew-Hessian}
  Let $m_c$ be a skew critical point of $(\alpha,\kD,g)$. Define the
  bilinear form $d_\kD\alpha(m_c)\colon
  \w{\!T\-4_}{m_c}M\times\kD_{m_c}\to\RR$ by
  \[
    \!d_\kD\alpha(m_c)(u,v)\=\bigl\langle\!d(\!i_V\alpha)(m_c),u\bigr\rangle,
  \]
  where $V$ is a (local) vector field with values in $\kD$ such that
  $V(m_c)=v$. The \defemph{skew Hessian} of $\alpha$ with respect to
  $g$ and $\kD$ is the bilinear form
  \[
    \!d_{\kD,g}\alpha(m_c)\colon\ker\w{\!T\-4_}{m_c}g\times\kD_{m_c}\to\RR
  \]
  obtained by restriction of $\!d_\kD\alpha(m_c)$. Define
  $\!d_{\kD,g}\alpha(m_c)^\flat\colon\ker\w{\!T\-4_}{m_c}g\to\kD_{m_c}^*$ by
  \[
    \!d_{\kD,g}\alpha(m_c)^\flat(u)\=\!d_{\kD,g}\alpha(m_c)(u,\,\cdot\,).
  \]
  A skew critical point $m_c$ of $(\alpha,\kD,g)$ is called
  \defemph{nondegenerate} if $\!d_{\kD,g}\alpha(m_c)^\flat$ is a
  linear isomorphism.
\end{definition}

We are using the term `skew critical problem' in two slightly
ways. First, it is any constrained critical point problem where the
derivative is taken in directions that may be other than the
constraint directions. Second, it is a problem within the technical
meaning of Definition~\ref{def:skew-critical-problem}. The two are not
really the same. For example, in
Definition~\ref{def:skew-critical-problem}, the purpose of the
function $g$ is to provide a parametrization of the critical points,
rather than a constraint. It is not necessarily a natural constraint
e.g.\ its level sets do not necessarily correspond to the leaves of
$\kD$, in the case that $\kD$ is integrable.

If $\kD_d$ is the level set of a submersive \defemph{constraint
function} $g_d\colon\kV\rightarrow\kP$ i.e.\ $\kD_d=g_d^{-1}(p_0)$,
then \LdApd\ may be cast as a skew critical problem within
Definition~\ref{def:skew-critical-problem}, in a variety of ways. One
of the ideas, though, of such skew critical problems, is that their
critical points should be isolated on the level sets of the constraint
function, and should be smoothly parametrized by values of the
constraint function. So, the constraint function of the skew critical
problem representing \LdApd\ should provide exactly the freedom
required to fit the skew critical points to the constraint levels and
to the initial conditions of the evolutions of \LdApd. Failure of
this would rule out (local) existence and uniqueness for discrete
evolutions as a consequence of the natural persistence of
(nondegenerate) skew critical points. The following is a formulation
of \LdApd\ as a skew critical problem in the atomic $N=2$ case; the
case for arbitrary $N$ is similar.

\begin{definition}
  The \defemph{associated skew critical problem} to the DCLS
  $(\kV,\w{\sigma\-6_d},\kD_d,\kE_d)$ is $(\Sigma_d,\kW_d,\hat g_d)$,
  defined on $\kC_d$, and where
  \[
    \hat g_d\colon\kC_d\rightarrow\kV\times\kP,\qquad
      \hat g_d(v,\tilde v)\=\bigl(v,g_d(\tilde v)\bigr).
  \]
\end{definition}

Clearly, $\hat g_d^{-1}(\hat p_0)\subseteq\kN_d$ for any choice $\hat
p_0\=(v,p_0)$ such that $g_d(v)=p_0$, and the solutions to
$(\Sigma_d,\kW_d,\hat g_d)$ as $v$ varies over $\kD_d$ are exactly the
solutions to \LdApd. The justification for the particular choice of
$\hat g_d$ comes from the following.

\begin{theorem}\label{thm:regular-DCLS-implies-nondegenerate-skew}
  All solution pairs of a regular$^+$ DCLS $(\kV,\w{\sigma\-6_d},\kD_d,\kE_d)$
  are nondegenerate skew critical points of the skew critical problem
  $(\Sigma_d,\kW_d,\hat g_d)$.
\end{theorem}

\begin{proof}
Let $(v_c,\tilde v_c)$ be a solution pair. It is required to prove
that $\w{\!T\-4_}{(v_c,\tilde v_c)}(\kV\times\kV)\ni(\delta v,\delta\tilde
v)=0$ if
\[[eq:regular-DCLS-implies-nondegenerate-skew-1]
  \left\{\begin{array}{l}
    \mbox{$(\delta v,\delta\tilde v)\in\ker\!d\hat g_d$, and}
    \\[2pt]
    \mbox{$\!d_{\kW_d,\hat g_d}\Sigma_d\bigl((\delta v,\delta\tilde v),
      (\delta w,\delta\tilde w)\bigr)=0$ for all
      $(\delta w,\delta\tilde w)\in(\kW_d)_{(v_c,\tilde v_c)}.$}
  \end{array}\right.
\]
Assuming~\eqref{eq:regular-DCLS-implies-nondegenerate-skew-1}, choose
$\delta q\in\kW_{\partial^+(v_c)}\kQ$, and extend $\delta q$ to a
(local) vector field $X_{\delta q}$ on $\kQ$, as in the development
just after Theorem~\ref{th:DCLS-fundamental-decomposition}. Then
$X\=(X_{\delta q}^+,X_{\delta q}^-)$ is a vector field in $\kW_d$, so
$(\delta w,\delta\tilde w)\in(\kW_d)_{(v_c,\tilde v_c)}$ if $\delta
w\=X^+_{\delta q}(v_c)$ and $\delta\tilde w\=X^-_{\delta q}(\tilde
v_c)$. Also,
\[[eq:half-differentiated-hessian]
  \!i_X\Sigma_d(v,\tilde v)=\!i_{X^+}\w{\sigma\-6_d}(v)
    +\!i_{X^-}\w{\sigma\-6_d}(\tilde v).
\]
$(\delta v,\delta\tilde v)\in\ker\!d\hat g_d$ implies, from the
definition of $\hat g_d$, that $\delta v=0$ and $\delta\tilde v\in
\w{\!T\-4_}{\tilde v_c}\kD_d$ where $\kD_d=g_d^{-1}(p_0)$ and $p_0\=g_d(v)$.
Since the skew critical problem occurs in $\kC_d$,
$\!T\partial^-(\delta\tilde v)=\!T\partial^+(\delta v)=0$, so it is
enough to show $\!T\partial^+(\delta\tilde v)=0$. The first term of
Equation~\eqref{eq:half-differentiated-hessian} differentiates to zero
in the direction $(\delta v,\delta\tilde v)$, and hence
\[
  0=\!d_{\kW_d}\Sigma_d\bigl((\delta v,\delta\tilde v),
    (\delta w,\delta\tilde w)\bigr)=\!d^\pm\w{\sigma\-6_d}(\tilde v_c) (\delta
    q,\delta\tilde q),
\]
where $\delta\tilde q=\!T\partial^+(\delta\tilde v)\in
\!T\partial^+(\w{\vrt^-_{\tilde v_c}\-3}\kV\cap\w{\!T\-4_}{\tilde
v_c}\kD_d)$. Then $\delta\tilde q=0$, since $\delta q$ is arbitrary
and the DCLS is  regular$^+$.\elsqed
\end{proof}

%
\section{Concluding remarks}
%
%
\noindent \emph{Discretization} is, in one view, the replacement of
infinitesimal objects with finite, geometric ones, depending on what
one wants to represent. In differential geometry this might mean the
attachment of geometric objects to every point of a manifold. The
ubiquitous linear bundles of differential geometry become bundles of
geometric shapes; the linear fibers become sets on which most of the
usual linear operations are absent. At $h=0$ there is degeneration.
Obtaining this limit requires desingularization, and, for coherent
results along the whole of a continuous target, semiglobal analysis.
For each $h$ of a discretization, we have a \emph{discrete analogue},
which is a simpler, abstract construct, because it has not the burden
of supporting the limit $h\to0$.

Using curve segments to discretize tangent bundles is geometrically
vivid. The abstract notion of a discrete tangent bundle
$(\kV,\partial^+,\partial^-)$
(Definition~\ref{def:discrete-tangent-bundle}) of a configuration
space $\kQ$, is an example of the clarity afforded by invariant
differential geometry:
$$
\includegraphics{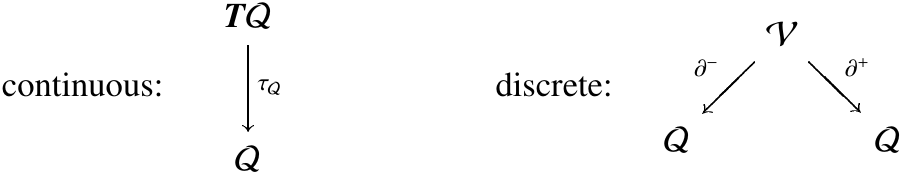}
$$
While they do not support linear operations, the discrete tangent
bundles have other structures. The forward and backward discrete
canonical projections result in decompositions because they split
every fiber of $\!T\kV$.

To work Hamilton's principle directly on velocity phase space is to
view the equation $q(t)^\prime=v$ as a constraint. The discrete
analogue of this constraint for a sequence $v_i\in\kV$ is that
successive curve segments attached to the $v_i$ join to make a
continuous whole i.e.\ $\partial^+(v_i)=\partial^-(v_{i+1})$. That,
together with the usual fixed endpoint constraints, and discrete
action the sum of a discrete Lagrangian over $v_i$, altogether define
the discrete variational principle that gives the discrete evolution.
What follows that is more-or-less a straight application of the
philosophy of~\cite{MarsdenJE-PatrickGW-ShkollerS-1998-1,
  PatrickGW-2007-1}, which extracts structural properties directly
from variational principles.

Practically, it is easier, and more direct, to generate curve segments
rather than interpolate between the configurations $(q^+,q^-)\in\kQ$.
In~\cite{PatrickGW-CuellC-SpiteriRJ-ZhangW2009-1} we derive, based on
the discretizations of this article, numerical methods for explicitly
constrained Lagrangian systems. The required curve segments may be
generated using virtually any one step numerical integrator, and,
automatically there follows a variational integrator of the same
order. The current state of the art in geometric integrators for
nonholonomic systems uses the MV discrete phase
space~\cite{CortesJ-MartinezS-2001-1,
  deLeonM-deDiegoD-SantamariaMerinoA-2004-1,
  FedorovYN-ZenkovDV-2005-1, McLachlanR-PerlmutterM-2006-1}. It is
future work to address construction of and the error analysis of
nonholonomic variational integration algorithms using the
discretizations of this article.

The curve segments naturally shrink to points as $h\to0$. This results
in a well defined and precise approach to the limit $h\to0$, of which
Proposition~\ref{prp:discretization-gives-discrete} shows typical use.
It is good to be respectful of this
limit. In~\cite{PatrickGW-CuellC-2008-1} we show that the error
analysis of discrete holonomic variational integrators, which is also
an issue of $h\to0$, depends on a subtle symmetry, and has sometimes
been oversimplified.

The discrete phase spaces and the continuous ones are conceptually
separated; there is no innate association between discrete and
continuous states, nor is there any unique configuration associated to
a discrete velocity. This is an unnatural conceptual point and it has
to be forcibly remembered, especially when, as is usual, the discrete
phase space and the continuous one are the same, set theoretically.
Since there is no innate association of continuous and discrete
states, there is neither any association of discrete states with
physical states. Such associations are inherently ambiguous if $h>0$.
\begin{figure}[h]
  \begin{center}\input{fig6_pstex_t.tex}\end{center}
\end{figure}

For example, when projecting to configuration space, there are not
just the two possibilities $\partial^+$ and $\partial^-$, but also
$\w{\tau\-2_\kQ}$, any other point on the curve segment, and, if
$\kV=\!T\kQ\subseteq\RR^N$, any convex combination of $\partial^+$ and
$\partial^-$. The continuous tangent vector associated to curve
segments is similarly ambiguous. Without further motivation, any of
these choices are as good as any others. With motivation, such choices
reflect the motivation, not the presence of a preeminent choice.
Suppose one has a variational integrator, and some association
$\!T\kQ\to\kV$. Conjugation by any structure preserving morphism,
which is near to the identity to sufficiently high order in $h$, gives
another association. The conjugated and original variational
integrators are equivalent; the implied change in association of
physical state to discrete states is not relevant.  And, in any case,
even though the discrete and continuous phase spaces may be the same
set theoretically, the structures of the discretization do not usually
have the same functional form as those of the continuous system. So
what is the justification for identifying the phase spaces?

If a discretization of a structured model is not structure preserving,
then it is subordinate and its states are slaved to continuous
states. If such a discretization is structure preserving and has equal
stature to a continuous model, then its states correspond to
continuous ones only ambiguously. Any specific identification of
continuous and discrete states, such as e.g.\ the identification of
the Lagrange multiplier of the discrete first order constraint with
the continuous momentum of the continuous Lagrangian, can only be
admitted a status similar to a \emph{possibly convenient special
coordinate system.} Of course many such coordinates exist in
geometry; but they are, at most, \emph{important and useful
intermediaries} which cannot properly be elevated to the stature of
necessity or structural centrality.

%% file: fig4_pstex_t.tex
\begin{picture}(0,0)%
\includegraphics{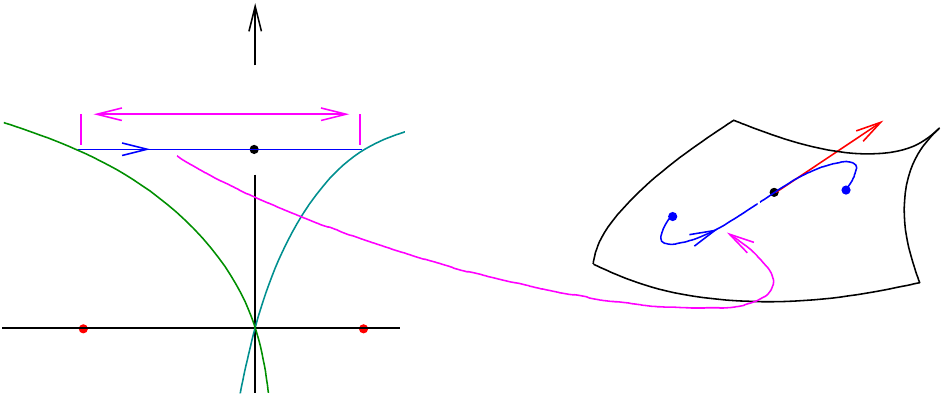}%
\end{picture}%
\setlength{\unitlength}{3947sp}%
\begingroup\makeatletter\ifx\SetFigFont\undefined%
\gdef\SetFigFont#1#2#3#4#5{%
  \reset@font\fontsize{#1}{#2pt}%
  \fontfamily{#3}\fontseries{#4}\fontshape{#5}%
  \selectfont}%
\fi\endgroup%
\begin{picture}(4522,1899)(316,-1638)
\put(758,-1460){\makebox(0,0)[b]{\smash{{\SetFigFont{8}{9.6}{\rmdefault}{\mddefault}{\updefault}$\alpha^-(h)$}}}}
\put(2123,-1460){\makebox(0,0)[b]{\smash{{\SetFigFont{8}{9.6}{\rmdefault}{\mddefault}{\updefault}$\alpha^+(h)$}}}}
\put(1386,-221){\makebox(0,0)[b]{\smash{{\SetFigFont{8}{9.6}{\rmdefault}{\mddefault}{\updefault}{\color[rgb]{1,0,1}$h+O(h^2)$}%
}}}}
\put(3146,-1521){\makebox(0,0)[lb]{\smash{{\SetFigFont{8}{9.6}{\rmdefault}{\mddefault}{\updefault}$t\mapsto\psi(h,t,m)$}}}}
\put(3690,-313){\makebox(0,0)[rb]{\smash{{\SetFigFont{8}{9.6}{\rmdefault}{\mddefault}{\updefault}$\kM$}}}}
\put(4455,-315){\makebox(0,0)[rb]{\smash{{\SetFigFont{8}{9.6}{\rmdefault}{\mddefault}{\updefault}$v_m$}}}}
\put(4109,-801){\makebox(0,0)[lb]{\smash{{\SetFigFont{8}{9.6}{\rmdefault}{\mddefault}{\updefault}$\partial_h^+(v_m)$}}}}
\put(3490,-676){\makebox(0,0)[lb]{\smash{{\SetFigFont{8}{9.6}{\rmdefault}{\mddefault}{\updefault}$\partial_h^-(v_m)$}}}}
\put(2501,-982){\makebox(0,0)[lb]{\smash{{\SetFigFont{8}{9.6}{\rmdefault}{\mddefault}{\updefault}$\psi$}}}}
\end{picture}%

%% file: f3_pstex_t.tex
\begin{picture}(0,0)%
\includegraphics{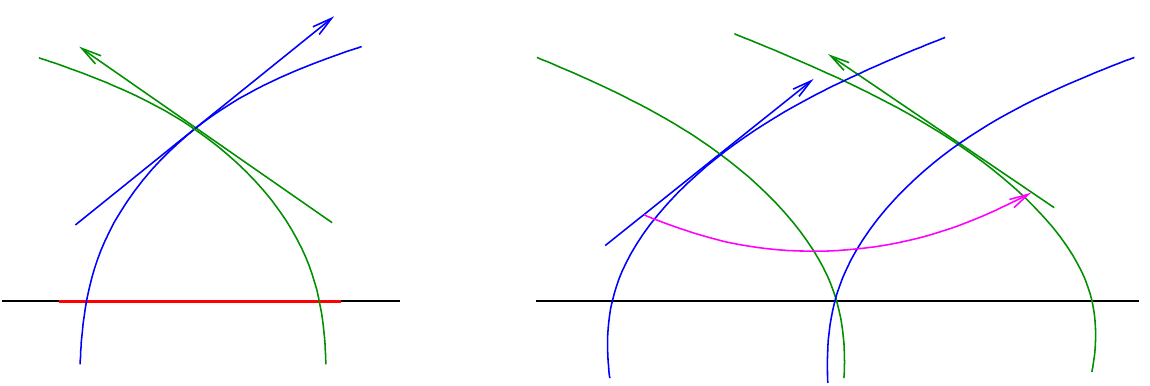}%
\end{picture}%
\setlength{\unitlength}{3947sp}%
\begingroup\makeatletter\ifx\SetFigFont\undefined%
\gdef\SetFigFont#1#2#3#4#5{%
  \reset@font\fontsize{#1}{#2pt}%
  \fontfamily{#3}\fontseries{#4}\fontshape{#5}%
  \selectfont}%
\fi\endgroup%
\begin{picture}(5503,1842)(701,-1419)
\put(5011,  6){\makebox(0,0)[lb]{\smash{{\SetFigFont{8}{9.6}{\rmdefault}{\mddefault}{\updefault}{\color[rgb]{0,.56,0}$\vrt^+_{\tilde v}\mskip-3mu\kV$}%
}}}}
\put(2166,300){\makebox(0,0)[rb]{\smash{{\SetFigFont{8}{9.6}{\rmdefault}{\mddefault}{\updefault}{\color[rgb]{0,0,1}$\vrt^-_v\mskip-3mu\kV$}%
}}}}
\put(1774,-211){\makebox(0,0)[lb]{\smash{{\SetFigFont{8}{9.6}{\rmdefault}{\mddefault}{\updefault}$v$}}}}
\put(4297,-346){\makebox(0,0)[lb]{\smash{{\SetFigFont{8}{9.6}{\rmdefault}{\mddefault}{\updefault}$v$}}}}
\put(5474,-291){\makebox(0,0)[lb]{\smash{{\SetFigFont{8}{9.6}{\rmdefault}{\mddefault}{\updefault}$\tilde v$}}}}
\put(2274,-946){\makebox(0,0)[lb]{\smash{{\SetFigFont{8}{9.6}{\rmdefault}{\mddefault}{\updefault}$\partial^+(v)$}}}}
\put(1104,-946){\makebox(0,0)[rb]{\smash{{\SetFigFont{8}{9.6}{\rmdefault}{\mddefault}{\updefault}$\partial^-(v)$}}}}
\put(4943,-871){\makebox(0,0)[lb]{\smash{{\SetFigFont{8}{9.6}{\rmdefault}{\mddefault}{\updefault}{\color[rgb]{1,0,1}$\tau_{\tilde v,v}$}%
}}}}
\put(2641,-1142){\makebox(0,0)[rb]{\smash{{\SetFigFont{8}{9.6}{\rmdefault}{\mddefault}{\updefault}$\kM$}}}}
\put(6189,-1142){\makebox(0,0)[rb]{\smash{{\SetFigFont{8}{9.6}{\rmdefault}{\mddefault}{\updefault}$\kM$}}}}
\put(1309, 94){\makebox(0,0)[lb]{\smash{{\SetFigFont{8}{9.6}{\rmdefault}{\mddefault}{\updefault}{\color[rgb]{0,.56,0}$\vrt^+_v\mskip-3mu\kV$}%
}}}}
\put(2281,  4){\makebox(0,0)[lb]{\smash{{\SetFigFont{8}{9.6}{\rmdefault}{\mddefault}{\updefault}{\color[rgb]{0,0,1}$\kV^-_{\mskip-2.5mu\partial^-(v)}$}%
}}}}
\put(1299,-76){\makebox(0,0)[rb]{\smash{{\SetFigFont{8}{9.6}{\rmdefault}{\mddefault}{\updefault}{\color[rgb]{0,.56,0}$\kV^+_{\mskip-2.5mu\partial^+(v)}$}%
}}}}
\put(4471, -7){\makebox(0,0)[rb]{\smash{{\SetFigFont{8}{9.6}{\rmdefault}{\mddefault}{\updefault}{\color[rgb]{0,0,1}$\vrt^-_v\mskip-3mu\kV$}%
}}}}
\end{picture}%

%% file: fig6_pstex_t.tex
\begin{picture}(0,0)%
\includegraphics{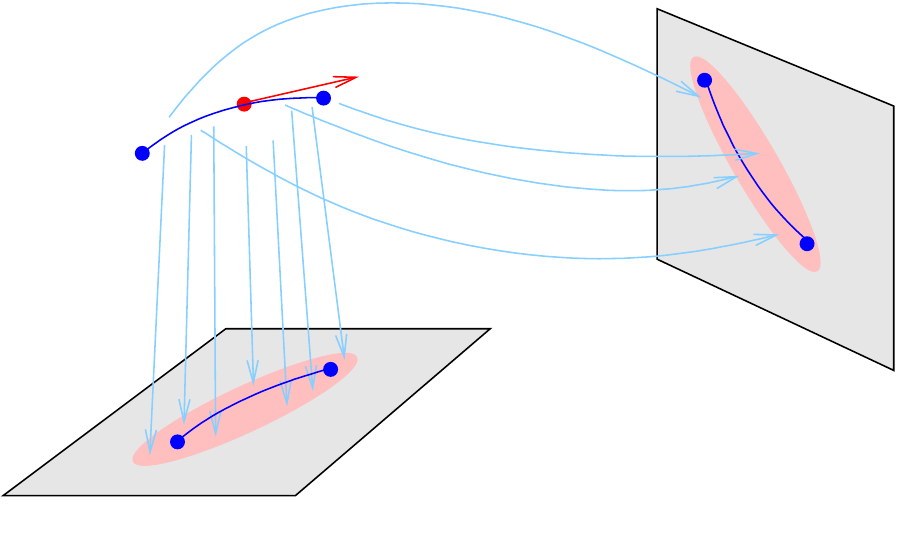}%
\end{picture}%
\setlength{\unitlength}{3947sp}%
\begingroup\makeatletter\ifx\SetFigFont\undefined%
\gdef\SetFigFont#1#2#3#4#5{%
  \reset@font\fontsize{#1}{#2pt}%
  \fontfamily{#3}\fontseries{#4}\fontshape{#5}%
  \selectfont}%
\fi\endgroup%
\begin{picture}(4302,2632)(1336,-3020)
\put(2723,-757){\makebox(0,0)[lb]{\smash{{\SetFigFont{8}{9.6}{\rmdefault}{\mddefault}{\updefault}{\color[rgb]{0,0,0}$v_q$}%
}}}}
\put(2464,-1006){\makebox(0,0)[lb]{\smash{{\SetFigFont{8}{9.6}{\rmdefault}{\mddefault}{\updefault}{\color[rgb]{0,0,0}$q$}%
}}}}
\put(1351,-2965){\makebox(0,0)[lb]{\smash{{\SetFigFont{8}{9.6}{\rmdefault}{\mddefault}{\updefault}{\color[rgb]{0,0,0}$\kQ$}%
}}}}
\put(5158,-628){\makebox(0,0)[lb]{\smash{{\SetFigFont{8}{9.6}{\rmdefault}{\mddefault}{\updefault}{\color[rgb]{0,0,0}$\!T\kQ$}%
}}}}
\end{picture}%